\documentclass[times,twocolumn,final]{elsarticle}
\usepackage{medima}
\usepackage{framed,multirow}

\usepackage{amssymb}
\usepackage{latexsym}

\usepackage{url}
\usepackage{xcolor}
\usepackage{color}
\usepackage{bm}
\usepackage{hyperref}
\usepackage{multicol}
\usepackage{multirow}
\usepackage{subcaption}
\usepackage{array}
\usepackage{verbatim}
\usepackage{booktabs}
\definecolor{newcolor}{rgb}{.8,.349,.1}

\usepackage{amsmath,amssymb,amsfonts}
\usepackage{algorithmic}
\usepackage{graphicx}
\usepackage{textcomp}
\usepackage[capitalise]{cleveref}
\Crefname{section}{Sec.}{Secs.}

\newcommand{\R}[1]{{\color{black} #1}}

\usepackage[T1]{fontenc}

\journal{Medical Image Analysis}

\begin{document}

\verso{Zhentao Liu \textit{et~al.}}

\begin{frontmatter}

\title{3D Vessel Reconstruction from Sparse-View Dynamic DSA Images via Vessel Probability Guided Attenuation Learning}

\author[1]{Zhentao \snm{Liu}}
\author[2]{Huangxuan \snm{Zhao}}
\author[1]{Wenhui \snm{Qin}}
\author[3]{Zhenghong \snm{Zhou}}
\author[3]{Xinggang \snm{Wang}}
\author[4]{Wenping \snm{Wang}}
\author[1]{Xiaochun \snm{Lai}}
\author[1]{Dinggang \snm{Shen}}
\author[1]{Zhiming \snm{Cui}}
\ead{cuizhm@shanghaitech.edu.cn}
%% Third author's email

\address[1]{School of Biomedical Engineering \& State Key Laboratory of Advanced Medical Materials and Devices, ShanghaiTech University, Shanghai, China}
\address[2]{National Engineering Research Center for Multimedia Software, School of Computer Science, Wuhan University, Wuhan, China}
\address[3]{School of Electronic Information and Communications, Huazhong University of Science and Technology, Wuhan, China}
\address[4]{Department of Computer Science \& Engineering, Texas A\&M University, USA}

% \received{1 May 2013}
% \finalform{10 May 2013}
% \accepted{13 May 2013}
% \availableonline{15 May 2013}
% \communicated{S. Sarkar}

\begin{abstract}
Digital Subtraction Angiography (DSA) is one of the gold standards for vascular disease diagnosis.
With the help of \R{a} contrast agent, time-resolved 2D DSA images deliver comprehensive blood flow information and can be utilized to reconstruct 3D vessel structures for medical assessment.
Current commercial DSA systems typically require hundreds of scanning views to perform reconstruction, resulting in substantial radiation exposure.
In this study, we propose a neural rendering-based optimization framework tailored for high-quality sparse-view DSA reconstruction to reduce radiation dosage.
Our approach, termed vessel probability guided attenuation learning, represents DSA imaging as a complementary weighted combination of static and dynamic attenuation fields, with the weights derived from the time-independent vessel probability field. 
Functioning as a \R{foreground} mask, vessel probability provides proper gradients for both static and dynamic fields adaptive to different scene types.
This mechanism enables self-supervised decomposition between static backgrounds and dynamic contrast agent flow, and significantly improves reconstruction quality.
Our model is trained by minimizing the discrepancy between synthesized projections and real captured DSA images.
We further employ two training strategies to improve reconstruction quality: (1) coarse-to-fine progressive training for better geometry and (2) temporal perturbed rendering loss for temporal consistency.
Experimental results have demonstrated high-quality 3D vessel reconstruction and 2D DSA image synthesis.
\end{abstract}


\end{frontmatter}

%\linenumbers
\section{Introduction}
\label{sec:introduction}

\begin{figure}[!t]
\centerline{\includegraphics[width=0.48\textwidth]{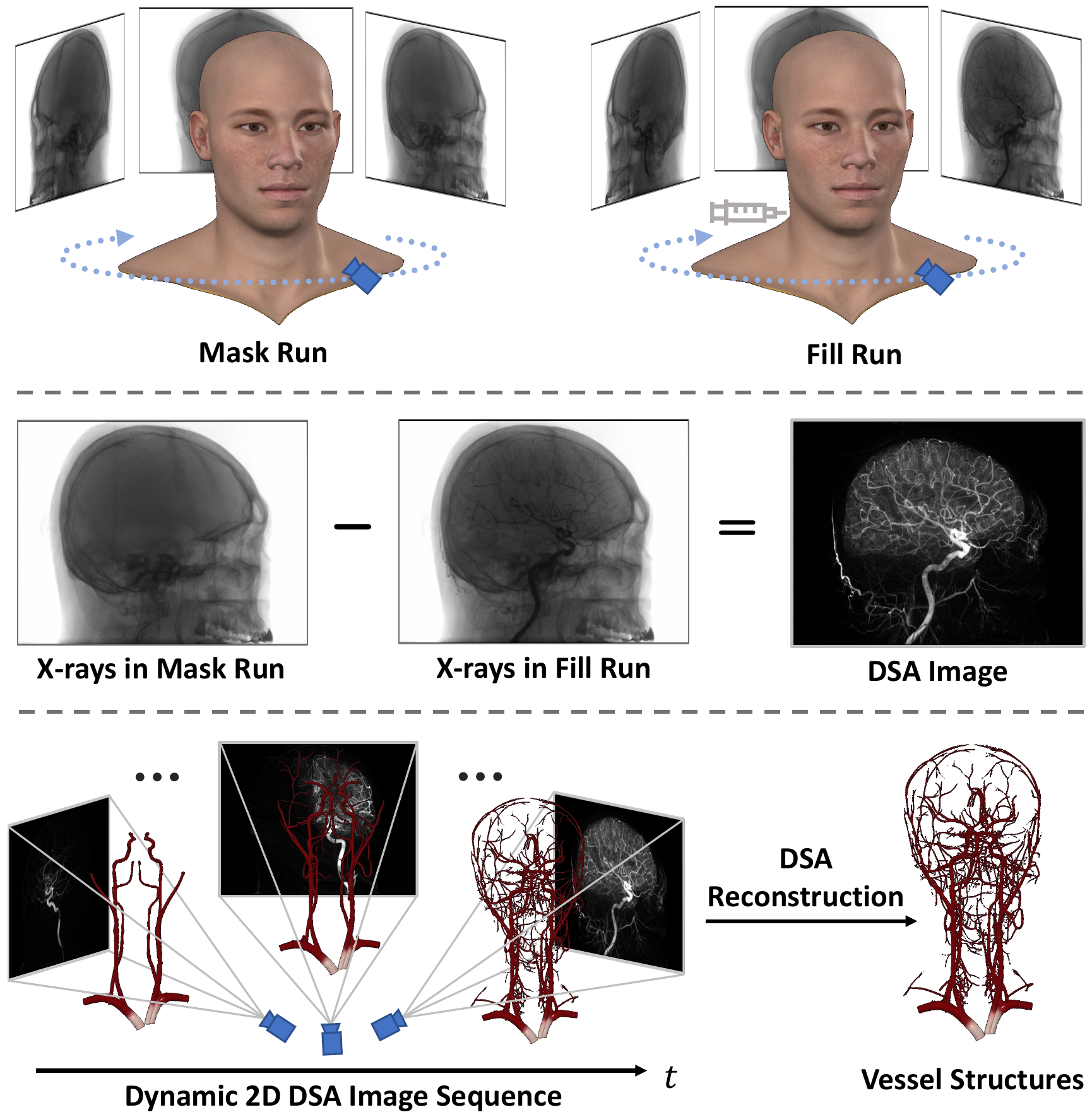}}
\caption{
DSA imaging involves two X-ray scans: a mask run and a fill run. 
Subtracting the fill run X-ray images from the mask run yields dynamic 2D DSA images, which are used to reconstruct 3D vessel structures.}
\label{fig:DSAimaging}
\vspace{-4mm}
\end{figure}

Digital Subtraction Angiography (DSA) is a crucial modality for diagnosing vascular diseases, such as stenosis, arteriovenous malformation (AVM), arteriovenous fistula (AVF), and intracranial aneurysms~\citep{DSA_app_1, DSA_app_3}.
As shown in \cref{fig:DSAimaging}, DSA imaging involves two rotational X-ray scans: a mask run acquired before contrast agent injection and a fill run acquired after. 
The DSA sequence is derived by subtracting the fill run X-ray images from those in \R{the} mask run.
This process highlights the blood flow information marked by contrast agent while removing non-vascular tissues.
Each DSA image captures a particular blood flow state as the contrast agent gradually flows through the vessels.
However, significant vessel overlap in DSA images would hinder accurate medical diagnosis.
To achieve a holistic understanding of vessel anatomy, the DSA sequence is then used to reconstruct 3D vessel structures.

In current commercial DSA systems, vessel reconstruction still relies on the traditional FDK (Feldkamp, Davis and Kress) algorithm~\citep{FDK,fdk-3ddsarecon}, typically requiring hundreds of scanning views to produce high quality reconstructions without artifacts.
Such a large number of X-ray projections induces significant radiation exposure for both patients and radiographers.
Thus, sparsifying scanned views to reduce radiation is pressingly needed.
Sparse-view DSA reconstruction poses a great challenge due to both dynamic imaging and data insufficiency.

With the advancement of deep learning, many attempts have been made to apply Convolutional Neural Networks (CNNs) and transformers for sparse-view Cone Beam Computed Tomography (CBCT)~\citep{Single-recon,X2CTGAN,DIFnet,3DCBCTrecon,X-LRM,X-GRM} and DSA reconstructions~\citep{3DDSArecon} in a feed-forward manner.
However, all these methods are designed for static volume reconstruction and are not well-suited for the dynamic DSA images collected in clinics.

Recently, Neural Radiance Field (NeRF)~\citep{NeRF} is becoming a powerful technique for high-quality novel view synthesis~\citep{DVGO, Instant-ngp, Freenerf} and surface reconstruction~\citep{neus,neuralangelo}.
It has been widely applied in static CBCT reconstruction and achieves promising results~\citep{Intratomo, NAF, sax_nerf}.
Notably, TiAVox~\citep{TiAVox} stands out as a relevant work, extending DVGO~\citep{DVGO} from temporal dimension to recover DSA scanning process.
However, its 4D voxel grid representation would lead to inefficient temporal modeling, resulting in noise and detail loss in reconstruction.

In this study, we propose a NeRF-based optimization framework tailored for sparse-view DSA reconstruction.
In fact, DSA imaging inherently possesses a strong structure prior knowledge: vessels remain \R{anatomically} static during \R{the} scanning process, serving as time-independent envelopes with dynamic contrast agent flowing through.
Based on this key observation, we propose a time-agnostic vessel probability field to capture DSA dynamic imaging nature effectively.
Our approach, termed vessel probability guided attenuation learning, represents DSA imaging as a complementary weighted combination of static and dynamic attenuation fields, with the weights derived from the vessel probability field.
A higher vessel probability emphasizes the dynamic field, indicating the presence of vessels and stronger dynamic contrast flow.
As a result, loss gradients primarily update the dynamic field's parameters, leaving the static field unchanged.
Conversely, a lower probability emphasizes the static field, corresponding to static background regions.
In this case, loss gradients chiefly modify the static field's parameters without disrupting the dynamic field.
Functioning as a foreground mask, vessel probability provides proper gradients for both static and dynamic fields adaptive to different scene types.
It enables self-supervised static-dynamic decomposition between static backgrounds and dynamic contrast agent flow, and significantly improves reconstruction quality.
We train our model by minimizing the discrepancy between synthesized projections and real captured DSA images.
Furthermore, we employ two training strategies to improve reconstruction quality: (1) coarse-to-fine progressive training to achieve better geometry, and (2) temporal perturbed rendering loss to enhance temporal consistency.
Our approach achieves high-quality 3D vessel reconstruction and 2D DSA image synthesis as demonstrated in experiments.

\section{Related Work}

\subsection{Traditional DSA Reconstruction}
\label{sec:relatedworks_traditional}

FDK algorithm~\citep{FDK,fdk-3ddsarecon} is the most widely used reconstruction method in DSA systems.
Modern systems have improved the original FDK, enabling high-quality vessel reconstruction from dynamic DSA images.
However, reliable performance typically requires hundreds of scanning views (e.g., 133 views in this study), resulting in substantial radiation exposure.
Additionally, the built-in FDK algorithm in DSA systems is often inaccessible in practice. 
Conventional FDK and other iterative CBCT reconstruction methods~\citep{SART,ASDPOCS} are designed for static CBCT scans and generally fail to handle the temporal dynamics present in DSA imaging. 
Furthermore, these methods tend to produce severe artifacts under sparse-view \R{conditions} due to insufficient measurements.

\subsection{Deep Learning-Based DSA Reconstruction}
\label{sec:relatedworks_DL}

With advances in deep learning, many methods have utilized the generalization ability of CNNs~\citep{Single-recon,X2CTGAN,DIFnet,3DCBCTrecon} to learn the mapping between X-ray projections and CBCT images.
Some of them~\citep{DIFnet,3DCBCTrecon} incorporate geometry-accurate feature querying aligned with CBCT scanning geometry, and achieve promising results even with extremely sparse inputs, such as 5 or 10 projections.
Zhao et al.~\citep{3DDSArecon} proposed a CNN-based framework for sparse-view DSA reconstruction using self-supervised projection loss.
It employs Digitally Reconstructed Radiograph (DRR)~\citep{drrref2} technique to simulate multi-view static projections from a well-reconstructed 3D vessel volume obtained via the DSA system, and then performs reconstruction.
However, this procedure is unsuitable for our dynamic DSA images.

Recent advances in computer vision have seen the rise of large-scale, feed-forward networks capable of rapid 3D reconstruction from sparse-view images within seconds~\citep{LRM,GRM,VGGT}.
Architectures like VGGT~\citep{VGGT}, a transformer-based model pretrained on extensive 3D datasets, can produce a full 3D reconstruction in a single forward pass.
Related feed-forward architectures have also been adapted for sparse-view CBCT reconstruction~\citep{X-LRM, X-GRM}.
\R{However, applying these data-driven approaches to DSA remains challenging due to the dynamic nature of DSA imaging and the scarcity of large-scale clinical datasets.}
% However, their potential for DSA reconstruction remains a promising but unexplored direction.

\subsection{Neural Rendering-Based DSA Reconstruction}
\label{sec:relatedworks_NeRF}

NeRF~\citep{NeRF} has rapidly progressed in the field of computer vision, particularly excelling in novel view synthesis~\citep{DVGO, Instant-ngp, Freenerf} and surface reconstruction~\citep{neus,neuralangelo}.
The core concept is to use differentiable volumetric rendering in combination with Implicit Neural Representation (INR) through Multi-layer Perceptrons (MLPs) to learn the density and radiance distribution of a 3D scene.

NeRF techniques have been widely applied in sparse-view CBCT reconstruction~\citep{Intratomo, NAF, sax_nerf}, achieving promising results.
Instead of learning a radiance field in natural scenes, they optimize an attenuation field to reconstruct CBCT image.
However, all of them are designed for static CBCT scans, making them unsuitable for dynamic DSA reconstruction.
TiAVox~\citep{TiAVox} stands out as a relevant work.
It extends DVGO~\citep{DVGO} in temporal dimension and uses learnable 4D voxel grids to recover the dynamic DSA imaging process.
However, its straightforward 4D representation does not consider the inter-frame relationships in DSA sequence, leading to inefficient temporal modeling.
In DSA imaging, most of the scene remains static, and thus directly storing static points in 4D voxel grids wastes model capacity as they occupy redundant voxels.
As a result, its reconstruction tends to suffer from noise and detail loss.

\subsection{\R{Dynamic NeRF and Scene Decomposition}}
\label{sec:relatedworks_dynamicNeRF}

\R{NeRF extensions for dynamic scene modeling can be broadly categorized into canonical-mapping methods and time-encoding methods.
Canonical-mapping approaches~\citep{D-nerf, nerfies, tineuvox}, exemplified by D-NeRF~\citep{D-nerf}, represent dynamic scenes by learning a deformation field that maps time-varying observations to a static canonical space.
While effective for smooth motion, these methods often struggle with complex topological or content changes.
HyperNeRF~\citep{hypernerf} alleviates this limitation by lifting the canonical space into a higher-dimensional hyper-space.}

\R{In contrast, time-encoding methods~\citep{kplanes, HexPlane, 4dhash} jointly encode spatiotemporal coordinates to model time-varying density and radiance. 
Although more flexible, such approaches can be under-constrained due to the high dimensionality of the solution space.
To improve modeling efficiency, dynamic scene decomposition techniques~\citep{Dynamic_nerf, masked_hash, Emernerf} explicitly separate static backgrounds from time-varying components, mitigating the adverse influence of static regions on the dynamics optimization.
For example, Gao et al.~\citep{Dynamic_nerf} rely on pre-segmented dynamic object masks during model optimization.
MSTH~\citep{masked_hash} introduces a learnable static mask to adaptively fuse 3D and 4D encodings.
EmerNeRF~\citep{Emernerf} decomposes the scene based on predicted static and dynamic densities.}

\R{These decomposition strategies provide important insights for our method.
In DSA imaging, only the contrast agent exhibits temporal variation within static vessel structures.
Accordingly, we introduce a time-invariant vessel probability field to explicitly model the static vascular anatomy.
This field serves as a foreground mask to guide gradient flow during optimization, thereby enabling effective static-dynamic decomposition and improving reconstruction quality.}

\section{Methods}
\label{sec:method}

In this section, we first introduce DSA imaging and reconstruction in \cref{sec:method preliminary}.
Next, we delve into our methodology, i.e., vessel probability guided attenuation learning, in \cref{sec:method VPAL}. 
Our training strategies, including coarse-to-fine progressive training and temporal perturbed rendering loss, are described in \cref{sec:method model optimization}.
Finally, vessel structures can be reconstructed by inferring our trained model in \cref{sec:method vessel reconstruction}.
A key notation table central to our method is given in \cref{table:notation definitions} for clarity.

\begin{table}[!t]
\centering
\caption{Key notation definitions.}
\label{table:notation definitions}
\fontsize{8}{8}\selectfont
\setlength{\tabcolsep}{2pt}
\renewcommand\arraystretch{1.2}
\begin{tabular}{ll}
\toprule
\textbf{Symbol} & \textbf{Description} \\
\midrule

$\bm{x}$          & 3D spatial coordinate.                                          \\
$t$                 & Continuous time variable.                                       \\
$I(\bm{r},t)$ & Ground truth DSA pixel value at ray $\bm{r}$ and time $t$. \\
$\hat{I}(\bm{r},t)$ & Synthesized DSA pixel value at ray $\bm{r}$ and time $t$.\\
$\mathcal{M}$ & Attenuation field, $\mathcal{M}:(\bm{x}, t) \in \mathbb{R}^3 \times \mathbb{R} \rightarrow \mu_c \in \mathbb{R}_{\geq0}$. \\
$\mu_c(\bm{x},t)$ & Predicted contrast agent attenuation value at $(\bm{x},t)$. \\
$\mathcal{S}$ & Static attenuation field, $\mathcal{S}:\bm{x} \in \mathbb{R}^3 \rightarrow \mu_s \in \mathbb{R}_{\geq0}$. \\
$\mu_s(\bm{x})$ & Predicted static attenuation value at point $\bm{x}$. \\
$\bm{h}_s$, $\phi_s$ & 3D hash encoder and decoding MLP that parameterize $\mathcal{S}$.\\
$\mathcal{D}$ & Dynamic attenuation field, $\mathcal{D}:(\bm{x}, t) \in \mathbb{R}^3 \times \mathbb{R} \rightarrow \mu_d \in \mathbb{R}_{\geq0}$. \\
$\mu_d(\bm{x},t)$ & Predicted dynamic attenuation value at $(\bm{x},t)$. \\
$\bm{h}_d$, $\phi_d$ & 4D hash encoder and decoding MLP that parameterize $\mathcal{D}$.\\
$\mathcal{P}$ & Vessel probability field, $\mathcal{P}:\bm{x} \in \mathbb{R}^3 \rightarrow p \in (0,1)$. \\
$p(\bm{x})$ & Predicted vessel probability value at point $\bm{x}$. \\
$\bm{h}_p$, $\phi_p$ & 3D hash encoder and decoding MLP that parameterize $\mathcal{P}$.\\
$\Delta t$ & Interval between neighboring training timestamps. \\
$\tau$ & Temporal perturbation as Gaussian noise, $\tau \sim \mathcal{N}(0, \sigma^2)$. \\
$k$ & Perturbation size that controls temporal smoothing strength. \\
$\sigma$ & Standard deviation of temporal perturbation, $\sigma=k \Delta t$. \\
$\mathcal{L}_1$ & L1 loss between synthesized and ground truth DSA pixel values. \\
$\mathcal{L}_{\text{reg}}$ & Regularization loss for vessel probability field. \\
$\lambda_{\text{reg}}$ & Weighting factor for regularization loss. \\
$\mathcal{L}$ & Total training loss. \\

\bottomrule
\end{tabular}
\vspace{-1mm}
\end{table}

\subsection{DSA Imaging and Reconstruction}
\label{sec:method preliminary}

\subsubsection{DSA Imaging Process}

As illustrated in \cref{fig:DSAimaging}, DSA imaging involves two rotational X-ray scans at the same position: the mask run, conducted before contrast agent injection, and the fill run, performed a few seconds after.
During both scans, the X-ray source rotates along a predefined arc-shaped trajectory while a 2D detector captures multi-view projections of the body (e.g., head and neck) at uniform angular intervals.
The primary difference between these two scans lies in the notably increased attenuation within vessels during the fill run due to the injected contrast agent.
Subsequently, the fill run X-ray images are subtracted from those in the mask run.
This process effectively removes non-vascular tissues, such as bones and muscles, resulting in DSA images that exclusively capture blood flow marked by contrast agent.

Each DSA frame captures both vascular morphology and time-resolved blood flow states as contrast agent flows through vessels.
Morphological features, such as vessel shape, percent stenosis~\citep{NASCET, WASID}, and aneurysm geometry~\citep{Demo_to_Neck}, can be directly observed or quantified from rotational DSA images.
These images also provide qualitative hemodynamic information, which could be further quantified into hemodynamic metrics by reconstructing a time-resolved contrast volume sequence, as validated in an in-silico simulation study~\citep{DSA_recon_metric}.
Together, the morphological and hemodynamic information offers clinical insights for vascular diseases diagnosis.

\subsubsection{DSA Imaging Formulation}

\begin{figure}
\centering
    \includegraphics[width=0.48\textwidth]{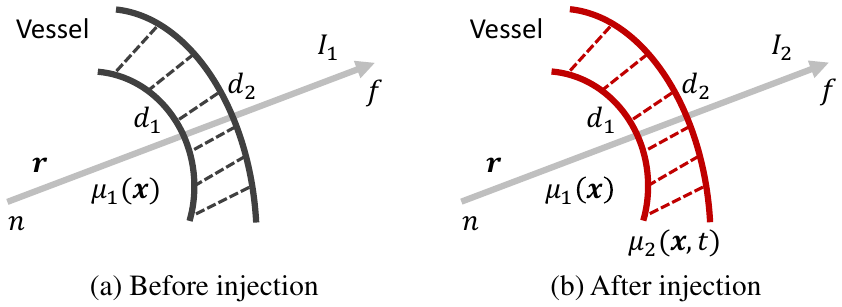}
    \caption{
    X-ray attenuation process along the X-ray path (a) before and (b) after contrast agent injection.
    } 
    \label{fig:dsaimaging_toyexp}
    \vspace{-4mm}
\end{figure}

\begin{figure*}[t!]
\centering
    \includegraphics[width=0.98\textwidth]{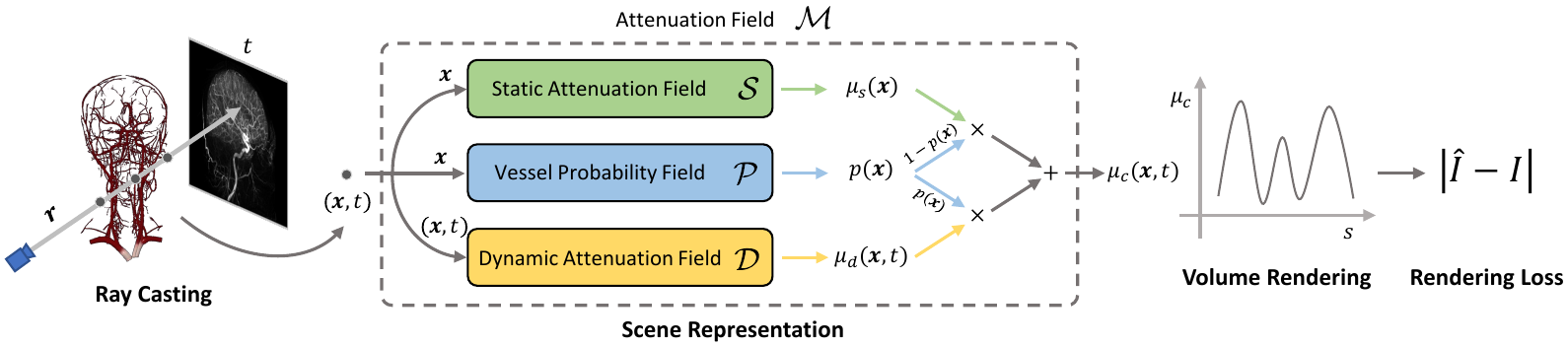}
    \caption{
    Overview of our proposed method.
    We represent the dynamic DSA imaging as a complementary weighted combination of static and dynamic attenuation fields, where the weights assigned to each component are derived from the vessel probability field.
    Our model is trained by minimizing the discrepancy between synthesized projections and real captured DSA images.
    } 
    \label{fig:method}
    \vspace{-4mm}
\end{figure*}

As illustrated in \cref{fig:dsaimaging_toyexp}, before contrast agent injection, the attenuation coefficient across the entire space is denoted as $\mu_{1}(\bm{x})\in\mathbb{R}_{\geq0}, \bm{x}\in\mathbb{R}^3$.
Consider an X-ray path given by $\bm{r}(s)=\bm{o}+s\bm{d} \in \mathbb{R}^3$, where $\bm{o}\in\mathbb{R}^3$, $s\in\mathbb{R}_{\geq0}$, and $\bm{d}\in\mathbb{R}^3$ are X-ray source position, length variable, and unit direction vector, respectively.
The X-ray attenuation process along this path is expressed via Beer's Law~\citep{beerslaw}:
\begin{equation}
    I_{1}(\bm{r}) = I_{0}(\bm{r})\exp{\left(-\int_{n}^{f}\mu_{1}(\bm{r}(s))\mathrm{d}s\right)},
\label{eq:beforeinjection}
\end{equation}
where $[n, f]$ is the path bound, $I_0(\bm{r})\in\mathbb{Z}_{\geq0}$ is the initial X-ray photon count emitted by the source, and $I_1(\bm{r})\in\mathbb{Z}_{\geq0}$ is the attenuated count measured by the detector.

After contrast agent injection, its local concentration within vessels changes over time as it circulates with blood flow, resulting in temporally varying attenuation in vessel areas $\mu_{2}(\bm{x},t)\in\mathbb{R}_{\geq0}, (\bm{x},t)\in\mathbb{R}^3\times\mathbb{R}$.
And the X-ray attenuation process along the same X-ray path transforms from \cref{eq:beforeinjection} to:
\begin{equation}
    I_{2}(\bm{r},t) = I_{1}(\bm{r})\exp{\left(-\int_{d_1}^{d_2}\left(\mu_{2}(\bm{r}(s),t)-\mu_{1}(\bm{r}(s))\right)\mathrm{d}s\right)},
\label{eq:afterinjection}
\end{equation}
where $[d_1, d_2]$ is the vessel segment on the X-ray path, and $I_2(\bm{r},t)\in\mathbb{Z}_{\geq0}$ is the attenuated photon count after injection.

Next, we subtract $I_2(\bm{r},t)$ from $I_1(\bm{r})$ in the logarithmic domain to get the DSA image pixel value $I(\bm{r},t)\in\mathbb{R}_{\geq0}$ as:
\begin{equation}
\begin{aligned}
    I(\bm{r},t) &= \ln{\left(I_1(\bm{r})\right)} - \ln{\left(I_2(\bm{r},t)\right)} \\
    &= \int_{d_1}^{d_2}\left(\mu_{2}(\bm{r}(s),t)-\mu_{1}(\bm{r}(s))\right)\mathrm{d}s \\
    &= \int_{n}^{f}\mu_{c}^{*}(\bm{r}(s),t)\mathrm{d}s,
\end{aligned}
\label{eq:dsarendering}
\end{equation}
where $\mu_{c}^{*}(\bm{x},t)\in\mathbb{R}_{\geq0}$ is the idealized target attenuation of contrast agent, which is defined as follows:
\begin{equation}
\mu_c^{*}(\bm{r}(s),t) = 
\begin{cases}
\mu_2(\bm{r}(s),t)-\mu_1(\bm{r}(s)) &   d_1 \leq s \leq d_2, \\
0 & \text{else}.
\end{cases}
\end{equation}
Note that $\mu_c^*$ is a conceptual target used for introducing DSA imaging and is not directly accessible for training supervision.

\subsubsection{Sparse-View DSA Reconstruction}

We denote 2D DSA image sequence as $\{\mathbf{I}_{i}\in \mathbb{R}^{w\times h}\}_{i=1}^{T}$, where $i$, $T$, and $w\times h$ are frame index, total number of frames, and image resolution, respectively.
The timestamp of the $i$-th frame is defined as $t_i=\frac{i}{T}\in(0,1]$, indicating its capture order.
The complete DSA frame data is then represented as $\{\mathbf{I}_{i}, t_{i}\}_{i=1}^{T}$.
Sparse-view DSA reconstruction aims to recover 3D attenuation volume representing vessel structures from a uniformly sampled subset $\{\mathbf{I}_{i_j}, t_{i_j}\}_{j=1}^{N}$, where $N<T$ and $i_j =\lfloor (j-1)\cdot\frac{T}{N} \rfloor+1$.

\subsection{Vessel Probability Guided Attenuation Learning}
\label{sec:method VPAL}

Our framework overview is depicted in \cref{fig:method}.
We aim to learn a 4D mapping function, termed attenuation field $\mathcal{M}$, that maps spatial-temporal coordinate $(\bm{x},t)$ to the contrast agent attenuation $\mu_c$, with $\mathcal{M}:(\bm{x}, t) \in \mathbb{R}^3 \times \mathbb{R} \rightarrow \mu_c \in \mathbb{R}_{\geq0}$.

\subsubsection{Naive Solution}
\label{sec:method naive solution}

In DSA imaging, most of the scene consists of static backgrounds, with only vessel areas containing flowing contrast agent.
Therefore, a straightforward idea is to decompose $\mathcal{M}$ into two components: the static attenuation field $\mathcal{S}:\bm{x} \in \mathbb{R}^3 \rightarrow \mu_s \in \mathbb{R}_{\geq0}$ and the dynamic attenuation field $\mathcal{D}:(\bm{x}, t) \in \mathbb{R}^3 \times \mathbb{R} \rightarrow \mu_d \in \mathbb{R}_{\geq0}$.
The former captures unchanging elements, while the latter depicts dynamic aspects.
The contrast attenuation is then expressed as their sum:
\begin{equation}
\mu_c(\bm{x}, t) = \mu_s(\bm{x}) + \mu_d(\bm{x}, t).
\label{eq:simpleadditioncomposition}
\end{equation}

Following Instant-NGP~\citep{Instant-ngp,4dhash}, $\mathcal{S}$ and $\mathcal{D}$ each comprise a multi-resolution hash encoder followed by a shallow MLP, with Rectified Linear Unit (ReLU) as the output activation. 
Thus, $\mu_s$ and $\mu_d$ are expressed as:
\begin{equation}
\mu_s(\bm{x}) =  \phi_s\left(\bm{h}_s(\bm{x})\right),~\mu_d(\bm{x},t) =  \phi_d\left(\bm{h}_d(\bm{x},t)\right),
\end{equation}
\R{where} $\bm{h}_s$ is a 3D hash encoder~\citep{Instant-ngp} to represent time-independent features, while $\bm{h}_d$ is a 4D hash encoder~\citep{4dhash} to capture time-varying features.
The following MLPs, denoted as $\phi_s$ and $\phi_d$, transform encoded features into attenuation values.
\cref{sec:method PG} provides more details about hash encoders.

We derive the gradients of $\mu_s$ and $\mu_d$ from \cref{eq:simpleadditioncomposition} as follows:
\begin{equation}
\frac{\partial \mu_c}{\partial \mu_s} = \frac{\partial \mu_c}{\partial \mu_d} = 1,
\label{eq:gradients_naive_solution}
\end{equation}
where $\mu_s$ and $\mu_d$ contribute equally to $\mu_c$.
However, this straightforward addition leads to poor reconstruction results, as model capacity is underutilized.
The majority of DSA imaging is time-invariant, such as static backgrounds, which could be accurately reconstructed by the static attenuation field alone.
Yet, these static points will always modify the dynamic attenuation field's parameters due to persistent gradients, as indicated by $\frac{\partial \mu_c}{\partial \mu_d} = 1$.
As a result, the dynamic field cannot accurately capture genuine dynamics.
Similarly, the static field will also be impaired by improper gradients, but not that severe, since dynamic flow is much sparser in the scene.
Thus, how to achieve proper gradients, especially that of dynamic field, is key to accurately represent dynamic DSA sequence.

\subsubsection{Vessel Probability Field}

To model the dynamic DSA imaging more effectively, we propose a time-agnostic vessel probability field $\mathcal{P}:\bm{x} \in \mathbb{R}^3 \rightarrow p \in (0,1)$.
It contains another 3D hash encoder $\bm{h}_p$ and decoding MLP $\phi_p$, with Sigmoid as the output activation. 
Thus, vessel probability $p$ is given by:
\begin{equation}
p(\bm{x}) = \phi_p(\bm{h}_p(\bm{x})).
\end{equation}

With the help of vessel probability, the attenuation expression transforms from \cref{eq:simpleadditioncomposition} to the following:
\begin{equation}
\mu_c(\bm{x}, t) = \left(1-p(\bm{x})\right)  \mu_s(\bm{x}) + p(\bm{x})  \mu_d(\bm{x}, t),
\label{eq:vpguidedcomposition}
\end{equation}
which represents $\mu_c$ as a complementary weighted combination of static attenuation $\mu_s$ and dynamic one $\mu_d$, with the weights determined by the vessel probability.

This design is based on the observation that contrast agents only flow within \R{anatomically} static vessels during the DSA scanning process.
We first derive the gradients of $p$ as follows:
\begin{equation}
    \frac{\partial \mu_c}{\partial p} = \mu_d - \mu_s.
\label{eq:gradients_of_vessel_probability}
\end{equation}
This equation tells us that $\mu_d$ yields positive gradients while $\mu_s$ contributes negative ones, causing them to compete with each other.
The final vessel probability map is shaped by the vessel areas where dynamic blood flows through, excluding the static backgrounds.
This results in an implicit capture of vessel topology by the vessel probability field.

Now the gradients of $\mu_s$ and $\mu_d$ change from \cref{eq:gradients_naive_solution} to:
\begin{equation}
\frac{\partial \mu_c}{\partial \mu_s} = 1-p,~\frac{\partial \mu_c}{\partial \mu_d} = p.
\label{eq:gradients_w_vessel_probability}
\end{equation}
A high value of $p(\bm{x})$ indicates vessel presence at point $\bm{x}$, typically associated with dynamic contrast flow.
It gives a higher weight to $\mu_d$ and a lower weight to $\mu_s$. 
During back-propagation, loss gradients mainly update the dynamic field's parameters, leaving the static field unaffected.
Conversely, a low value of $p(\bm{x})$ implies static backgrounds, increasing the contribution of $\mu_s$ while suppressing that of $\mu_d$.
In this case, loss gradients chiefly modify the static field's parameters without disrupting the dynamic field.
In summary, the vessel probability field provides proper gradients to both static and dynamic fields adaptive to different scene types.
Thus, static backgrounds and dynamic contrast agent flow could be accurately captured by static and dynamic fields without mutual interference.
Functioning as a foreground mask, vessel probability enables self-supervised static-dynamic decomposition and significantly improves reconstruction quality.
So far, we have solved the issue introduced in \cref{eq:simpleadditioncomposition}.

\subsection{Model Optimization}
\label{sec:method model optimization}

\subsubsection{Coarse-to-Fine Progressive Training}
\label{sec:method PG}

\begin{figure}
\centering
    \includegraphics[width=0.48\textwidth]{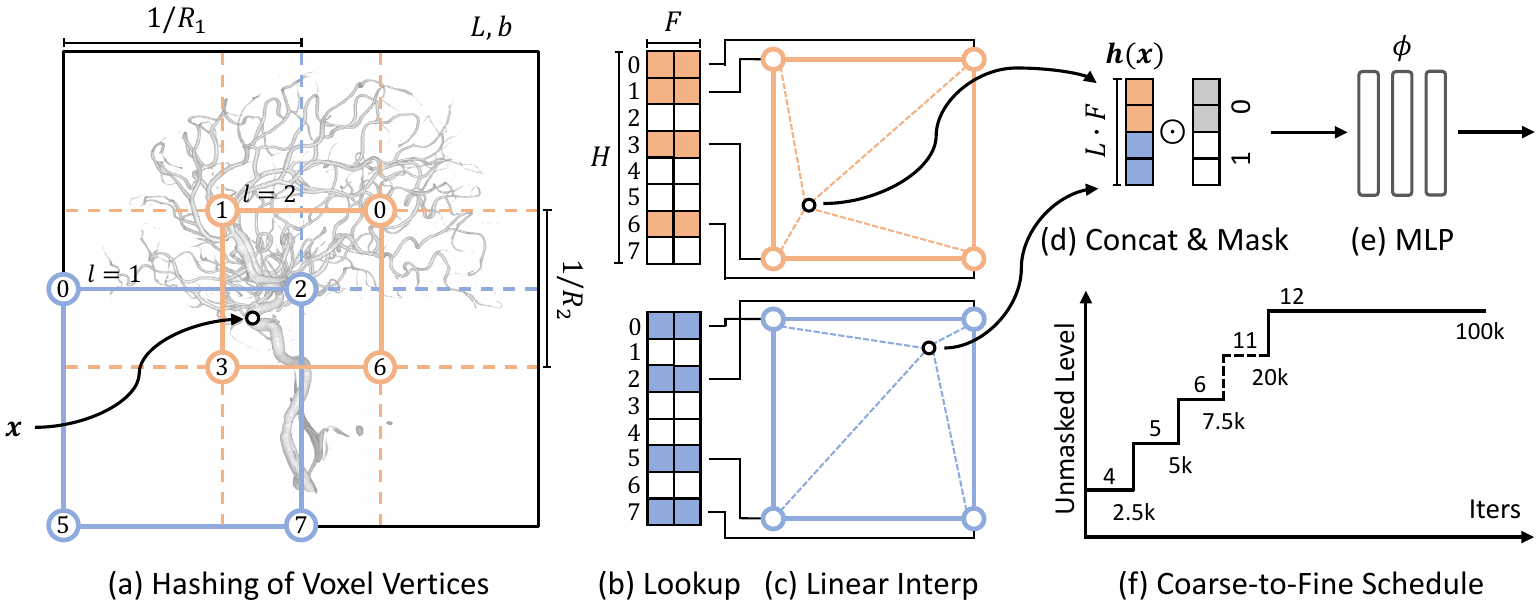}
    \caption{
    A 2D illustration of multi-resolution hash encoding with coarse-to-fine progressive training.
    (a) For input $\bm{x}$, vertices of its surrounding voxel at the $l$-th level are hashed to table indices, which are used to (b) look up $F$-dimensional features from a hash table of size $H$.
    (c) Features are linearly interpolated by the relative position of $\bm{x}$ within the $l$-th voxel.
    (d) All $L$ levels' interpolated features are concatenated and selectively activated via masking, as determined by (f) the coarse-to-fine schedule.
    (e) The final feature vector is decoded by a small MLP $\phi$.
    } 
    \label{fig:PG}
    \vspace{-4mm}
\end{figure}

As introduced in \cref{sec:method VPAL}, each of our three neural fields ($\mathcal{S},\mathcal{D},\mathcal{P}$) is parameterized by a multi-resolution hash encoder ($\bm{h}_s,\bm{h}_d,\bm{h}_p$) followed by a small MLP decoder ($\phi_s,\phi_d,\phi_p$).
We illustrate this process in \cref{fig:PG} with a 2D example.
The hash encoder represents the scene with $L$ levels voxel grids, where each level $l$ has resolution $R_{l}=\lfloor R_1 \cdot b^{l-1}\rfloor$, with $R_1$ as the coarsest resolution and $b$ the growth factor.
For input $\bm{x}$, the vertices of its surrounding voxel at the $l$-th level are hashed to table indices, which are used to look up $F$-dimensional trainable feature vectors from a hash table of size $H$.
These features are linearly interpolated based on the relative position of $\bm{x}$ within the $l$-th voxel, yielding a vector $\bm{h}^{l}(\bm{x})\in\mathbb{R}^F$.
The vectors from all levels are concatenated into $\bm{h}(\bm{x})= \left\{ \bm{h}^l(\bm{x}) \right\}^L_{l=1}\in\mathbb{R}^{L\cdot F}$ and fed into the MLP $\phi$ for decoding.

Hash grids at different resolutions capture scene details at varying scales: high-resolution grids encode fine details, while low-resolution ones capture coarse structures.
However, enabling all hash grids from the start of training can easily lead to high-frequency overfitting, hindering the model from exploring low-frequency information and causing undesired noisy artifacts~\citep{Freenerf}.
To mitigate this issue, we adopt a progressive training strategy~\citep{Freenerf,neuralangelo} that gradually unmasks different levels of hash grids from coarse to fine.
As shown in \cref{fig:PG}(d,f), we begin training with only the first few coarse levels ($l\leq4$) active by masking out feature vectors from finer grids.
As training progresses, higher-resolution levels are sequentially unmasked at predefined steps (e.g., every 2500 iterations).
This coarse-to-fine strategy stabilizes optimization by first capturing low-frequency components before fitting high-frequency details, thereby reducing noise and producing better vessel geometries with smoother surfaces.

As detailed in \cref{sec:method VPAL}, this entire pipeline is instantiated separately for the three neural fields to predict static attenuation $\mu_s$, dynamic attenuation $\mu_d$, and vessel probability $p$.

\subsubsection{Temporal Perturbed Rendering Loss}
\label{sec:method TP}

\begin{figure}
\centering
    \includegraphics[width=0.48\textwidth]{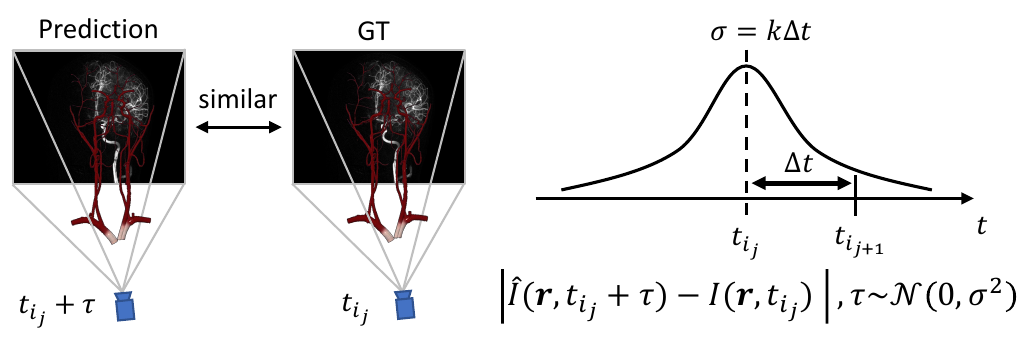}
    \caption{
    Illustration of temporal perturbed rendering loss. 
    Projection from the same viewpoint should be quite similar at slightly perturbed timestamps.
    The temporal perturbation is modeled as Gaussian noise.
    } 
    \label{fig:TP}
    \vspace{-4mm}
\end{figure}

For a ray $\bm{r}$ sampled from the $j$-th training frame, the model synthesizes pixel value $\hat{I}(\bm{r},t_{i_j})$ via the volumetric rendering defined in \cref{eq:dsarendering}.
\begin{equation}
\hat{I}(\bm{r},t_{i_j}) = \int_{n}^{f}\mu_c(\bm{r}(s),t_{i_j})\mathrm{d}s.
\label{eq:modelrendering}
\end{equation}
Our main training objective is to minimize L1 loss between synthesized pixel value and the ground truth $I(\bm{r},t_{i_j})$ over a batch of sampled rays.
\begin{equation}
\mathcal{L}_{1} = \mathbb{E}_{\bm{r},j} \left[ \left| \hat{I}(\bm{r}, t_{i_j}) - I(\bm{r}, t_{i_j}) \right| \right].
\label{eq:naive rendering loss}
\end{equation}

However, given sparse dynamic DSA images as training data, the model would overfit to the training frames without temporal regularization, producing temporally discontinuous floating artifacts.
A crucial insight is that blood flow is continuous over time, and the flow states between neighboring timestamps are expected to be quite similar.
Therefore, projections from the same viewpoint should remain stable at slightly perturbed timestamps.
Based on this observation, we propose a temporal perturbed rendering loss to enhance temporal consistency, extending the original loss in \cref{eq:naive rendering loss}:
\begin{equation}
    \mathcal{L}_1=\mathbb{E}_{\bm{r},j,\tau} \left[ \left| \hat{I}(\bm{r}, t_{i_j}+\tau) - I(\bm{r}, t_{i_j}) \right| \right], ~\tau \sim \mathcal{N}(0, \sigma^2).
    \label{eq:perturbed rendering loss}
\end{equation}

Essentially, this strategy acts as a temporal low-pass filter (\cref{fig:TP}).
The interval between neighboring training timestamps is $\Delta t = t_{i_{j+1}}-t_{i_j}$.
Temporal perturbation is modeled as Gaussian noise, $\tau \sim \mathcal{N}(0, \sigma^2)$ with standard deviation $\sigma$.
We set $\sigma=k \Delta t$, where $k\in\mathbb{R}_{\geq0}$ is the perturbation size controlling temporal smoothing strength.
This technique encourages the model to learn temporally smooth contrast dynamics, thereby suppressing discontinuous floating artifacts.

\subsubsection{Vessel Probability Regularization}

Furthermore, the vessel probability field is inherently under-constrained.
Given that vessel areas are sparse in DSA imaging scenes, we aim to drive the vessel probability generally towards zero, appearing only in necessary locations. 
Thus, we introduce another regularization term to encourage the sparsity of vessel probability:
\begin{equation}
\mathcal{L}_{\text{reg}} = \mathbb{E}_{\bm{x}} \left[ p(\bm{x}) \right].
\end{equation}
This loss is calculated from a batch of points $\bm{x}$ randomly sampled within the scanning bounding box.

% Compared to several alternatives, our L1 regularization is easy to implement without hyperparameter tuning, effectively encourages probability sparsity, and preserves fine vessel details.
Compared to several alternatives, our L1 regularization is easy to implement \R{and effectively encourages probability sparsity}.
Total Variation (TV) loss enhances spatial smoothness, but it is redundant as our MLP-based neural fields are inherently smooth~\citep{mlpbias}, and it does not promote sparsity.
Entropy-based loss encourages a bimodal (0 or 1) distribution, but its symmetry may suppress faint vessel details in highly imbalanced DSA data.
Alleviating this with focal loss~\citep{focalloss} introduces extra hyperparameter tuning.

\subsubsection{Overall Training Objective}

Our overall loss is a weighted sum of $\mathcal{L}_1$ and $\mathcal{L}_{\text{reg}}$, with $\lambda_{\text{reg}}$ adjusting regularization impact:
\begin{equation}
\mathcal{L} = \mathcal{L}_1 + \lambda_{\text{reg}} \mathcal{L}_{\text{reg}}.
\end{equation}

\subsection{Vessel Reconstruction}
\label{sec:method vessel reconstruction}

Once the model is well-trained, we can reconstruct the contrast agent attenuation volume at any time $t$ by querying the attenuation field $\mathcal{M}$ on voxel grids: $\mathbf{V}_c(t) = \left\{ \mu_c(\bm{x}, t) \mid \bm{x} \in \mathbb{X} \right\}\in \mathbb{R}^{W \times H \times D}$. 
Here, $\mathbb{X}\subset\mathbb{R}^3$ is the set of voxel center coordinates defined by the target volume's resolution $W\times H\times D$ and spacing.
The final 3D vessel volume is then obtained by averaging these volumes across all timestamps $\{ t_i \}_{i=1}^{T}$: $\overline{\mathbf{V}}_c = \frac{1}{T}\sum_{i=1}^{T}\mathbf{V}_c(t_i)$.
Similarly, we derive the vessel probability volume $\mathbf{V}_p$, the static attenuation volume $\mathbf{V}_s$, and the dynamic attenuation volume $\mathbf{V}_d(t)$ from their respective fields $\mathcal{P}$, $\mathcal{S}$, and $\mathcal{D}$, all of which lie in $\mathbb{R}^{W \times H \times D}$.
The averaged dynamic attenuation volume is computed as $\overline{\mathbf{V}}_d = \frac{1}{T}\sum_{i=1}^{T}\mathbf{V}_d(t_i)$.
These outputs can be combined to analyze different components of contrast agent.
Specifically, we compute the static component $(\mathbf{1}-\mathbf{V}_p)\mathbf{V}_s$, the dynamic component $\mathbf{V}_p\mathbf{V}_d(t)$, and the averaged dynamic component $\mathbf{V}_p\overline{\mathbf{V}}_d$.
Here, $\mathbf{1} \in \mathbb{R}^{W \times H \times D}$ is the \R{all-ones volume}.
For more details and visualizations, please refer to \cref{sec:model output}.

\section{Experiments}

\subsection{Experimental Settings}

\subsubsection{Dataset}
\label{sec:exp dataset}

\begin{table}[t]
\caption{Details of DSA images and reconstructed volumes.}
\label{table:data detail}
\centering
\fontsize{6.5}{6.5}\selectfont
\setlength{\tabcolsep}{3.5pt}
\renewcommand\arraystretch{1.2}
\begin{tabular}{ccccc}
\toprule
Cases \# & Image Resolution & Pixel Size ($\mathrm{mm}$) & Volume Resolution & Voxel Size ($\mathrm{mm}$) \\
\midrule
\R{7, 13, 19,} & \multirow{2}{*}{960$\times$960} & \multirow{2}{*}{0.3239$\times$0.3201} & \multirow{2}{*}{512$\times$512$\times$506} & \multirow{2}{*}{0.3802$^3$} \\
\R{24, 28, 29} & & & & \\
\cmidrule{1-5} % 中间加条细线区分
Others & 1240$\times$960 & 0.3219$\times$0.3208 & 512$\times$512$\times$395 & 0.4881$^3$ \\ 
\bottomrule
\end{tabular}
\vspace{-2mm}
\end{table}

\begin{table}[t]
\caption{List of hash encoder hyperparameters.}
\centering
\fontsize{6.5}{6.5}\selectfont
\setlength{\tabcolsep}{12.5pt}
\renewcommand\arraystretch{0.5}
\begin{tabular}{ccccccc}
\toprule
                     & $L$      & $H$ & $F$ & $b$  & $R_{1}$ & $R_{L}$ \\ \midrule
$\bm{h}_s, \bm{h}_p$ & 12 & $2^{19}$  & 8   & 1.45 & 8                & 476              \\
$\bm{h}_d$           & 12 & $2^{19}$ & 8   & 1.4  & 2                & 80               \\ \bottomrule
\end{tabular}
\label{table:hash_params}
\vspace{-4mm}
\end{table}

\R{\textbf{Dataset Descriptions.} In this study, we utilized 30 real-world patient cases acquired from two independent clinical centers, including 15 cases from Wuhan Union Hospital (WUH) and 15 from Renmin Hospital of Wuhan University (RHWU). 
All data were collected using the Siemens AXIOM-Artis DSA scanning system.
Cases are indexed as 1--15 (WUH) and 16--30 (RHWU) for clarity.}
The source-to-object and source-to-detector distances are 750$\mathrm{mm}$ and 1200$\mathrm{mm}$, respectively.
For each patient, the DSA system recorded 133 mask-fill X-ray image pairs during the cerebral arterial phase over a 198 degrees rotational range, with evenly distributed projection angles.
Additionally, vessel volumes reconstructed by the built-in FDK algorithm were provided by the system.
Although these volumes are not fully accurate and have imperfections (\cref{fig:imperfect_reference}), we use them as reference to evaluate 3D vessel reconstruction.
Details of the DSA images and reconstruction volumes are summarized in \cref{table:data detail}.
We uniformly subsampled 30, 40, 50, 60 views from the complete set as four training settings, with the rest used for 2D DSA image synthesis evaluation.

\R{\textbf{Ethics Statement.} This study was conducted in accordance with the Declaration of Helsinki and was approved by the Institutional Review Boards (IRBs) of Wuhan Union Hospital and Renmin Hospital of Wuhan University.
Due to the retrospective nature of the study, the requirement for informed consent was waived.
All patient data were fully anonymized and de-identified prior to analysis.}

\R{\textbf{Data and Code Availability.} The source code and a subset of anonymized test data are available at \url{https://github.com/ShanghaiTech-IMPACT/VPAL}.
We will further release the complete dataset once the administrative authorization for public sharing is granted.}

\subsubsection{Implementation Details}

Our proposed method comprises three implicit fields, each with a hash encoder followed by a three-layer MLP of hidden size 128.
\cref{table:hash_params} lists hyperparameters of three hash encoders ($\bm{h}_s,\bm{h}_d,\bm{h}_p$) used in our model.
These include \R{the} number of levels $L$, hash table size $H$, feature dimensions $F$, growth factor $b$, the coarsest ($R_{1}$) and finest ($R_{L}$) hash grids resolution.
For all three encoders, the initial active hash level is 4, and subsequent levels are activated every 2500 iterations.
We set regularization weight $\lambda_{\text{reg}}$ to 0.01, perturbation size $k$ to 1, random ray batch size to 2048, and regularization points batch size to 10k.
The Adam optimizer is employed for optimization, starting with an initial learning rate of $7.5\times10^{-4}$, which decays by a factor of 0.9 every 5k iterations. 
We train our model for a total of 100k iterations. 
All experiments are conducted on a single A100 GPU.

\subsubsection{Competing Methods}

We compare our method against the traditional FDK algorithm with hann filtering~\citep{FDK,fdk-3ddsarecon}, NeRF-based approaches such as NAF~\citep{NAF} and TiAVox~\citep{TiAVox}.
FDK is based on the TIGRE toolbox~\citep{tigre}, while other methods are adapted from their official source codes.
\R{Note that our method focuses on the per-case optimization setting.
Feed-forward reconstruction networks are excluded from comparison due to fundamental methodological differences, the lack of established baselines for dynamic DSA, and the scarcity of large-scale clinical datasets for training.}

\subsubsection{Evaluation Metrics}

\begin{figure}[t]
\centering
    \includegraphics[width=0.48\textwidth]{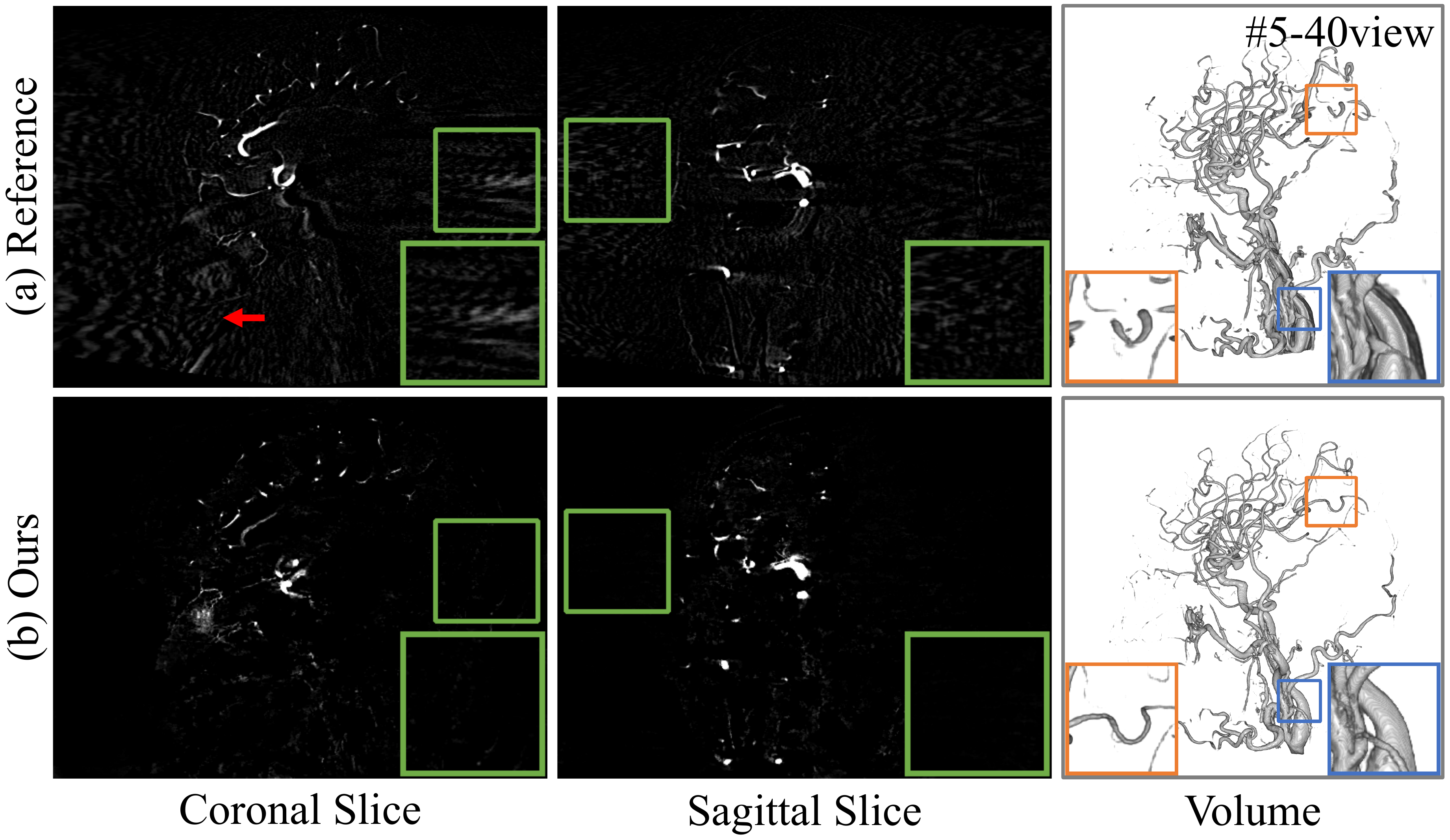}
    \caption{
    Reference volume imperfections.
    (a) Reference volume shows noise and streak artifacts (coronal and sagittal slices), broken vessels (orange box), and noisy surfaces (blue box);
    (b) our reconstruction mitigates such noise, and preserves fine vessel details with smoother surfaces.
    } 
\label{fig:imperfect_reference}
\vspace{-4mm}
\end{figure}

% \begin{figure}[t]
% \centering
%     \includegraphics[width=0.48\textwidth]{Img/max_intensity_comparison.pdf}
%     \caption{
%     Maximum intensity statistics analysis (mean$\pm$std) of our 15 cases.
%     The gray dashed line indicates that the voxel intensity scale of reference volume is approximately $4\times$ larger than that of reconstructions from different algorithms under varying input view settings.
%     } 
% \label{fig:max_intensity_comparison}
% \vspace{-4mm}
% \end{figure}

\begin{figure}[t]
\centering
    \includegraphics[width=0.49\textwidth]{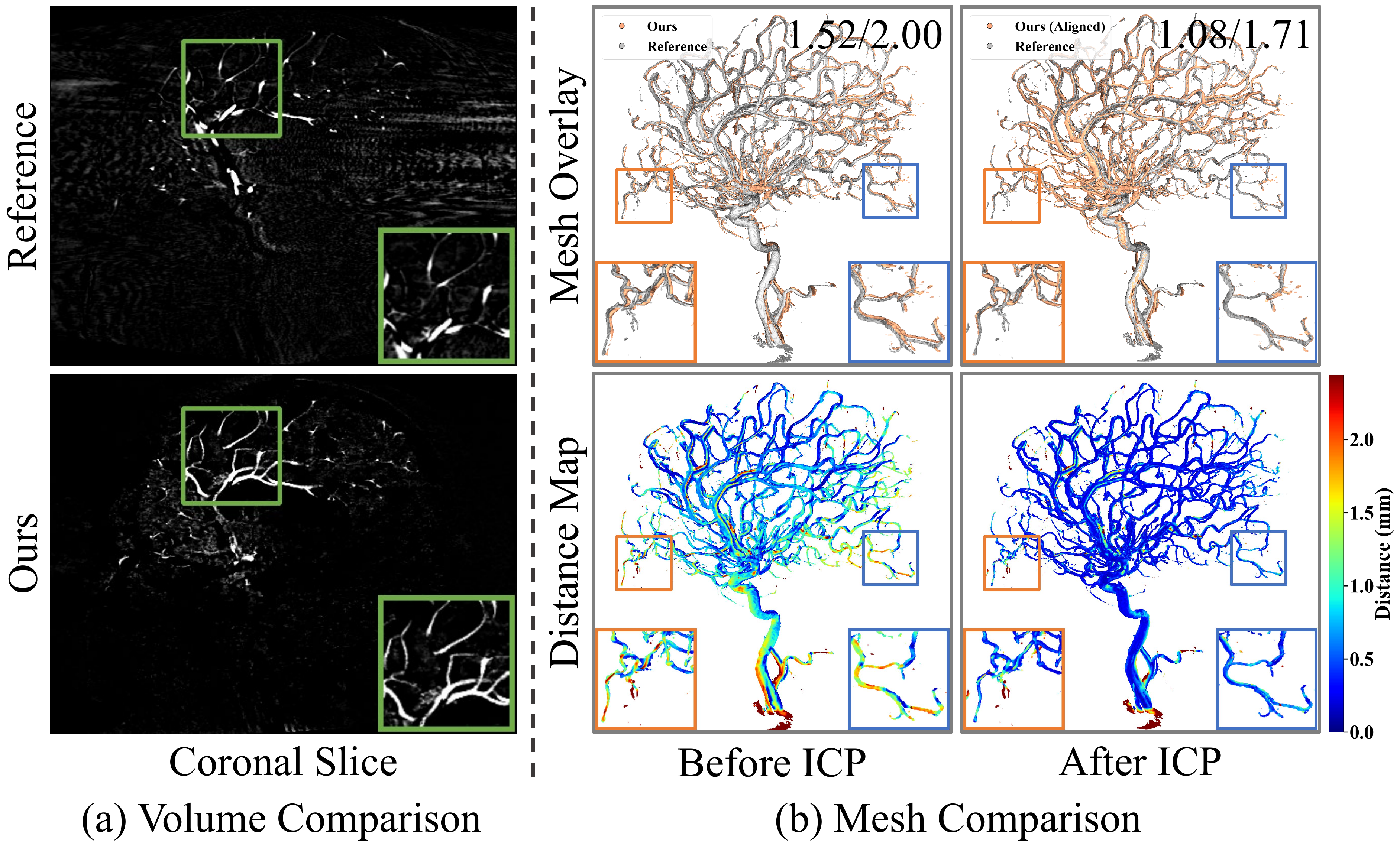}
    \caption{
    Illustration of spatial misalignment between reference volume and our method reconstruction, and its correction via the ICP algorithm.
    (a) Different content from the same coronal slice reveals a clear spatial shift.
    (b) Before alignment, this shift leads to large mesh surface distances as evidenced by distance map and CD($\mathrm{mm}$)/HD95($\mathrm{mm}$) metrics.
    After ICP, the reconstruction mesh aligns much closer to the reference with reduced surface distances.
    Results from case \#1 with 30 training views.
    } 
\label{fig:why_mesh}
\vspace{-4mm}
\end{figure}

\begin{figure}[t]
\centering
    \includegraphics[width=0.48\textwidth]{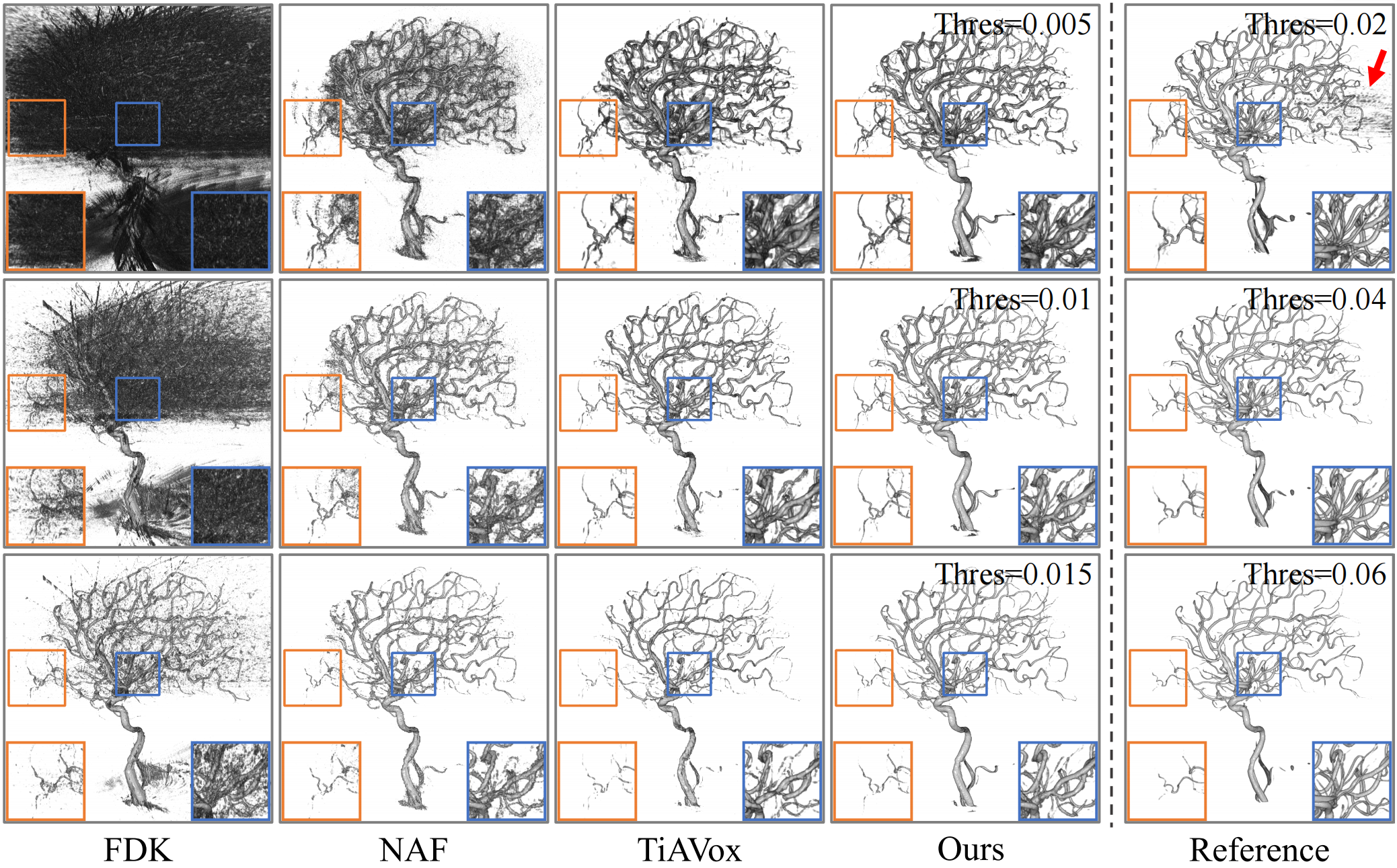}
    \caption{
    \R{Sensitivity analysis of Marching Cubes threshold selection} for reference and reconstructed volumes, with threshold shown top-right.
    Thresholds of 0.04 for reference and 0.01 for reconstructions achieve the best trade-off between noise suppression and detail preservation.
    Results from case \#1 with 30 views.
    } 
\label{fig:thres_justify}
\vspace{-4mm}
\end{figure}

We evaluate 3D vessel reconstruction (major task) and 2D DSA image synthesis (minor task).
Since FDK and NAF are static methods, we directly evaluate their reconstructed volumes.
TiAVox and our method are dynamic, so we compute the time-averaged reconstructions as the final output, as described in \cref{sec:method vessel reconstruction}.
Direct volumetric comparison between algorithm reconstructions and reference volume is unreliable due to both geometry and intensity calibration issues.
First, a spatial misalignment exists, likely caused by minor discrepancies between the scanner's proprietary FDK geometry and our DICOM-derived geometry used by evaluated algorithms.
This is evident by slight shift observed at both volume and mesh levels in \cref{fig:why_mesh} (with mesh extraction detailed later).
Second, the reference volume's intensity scale is roughly $4\times$ higher than that of reconstructions, and their voxel values have distinct physical meanings: the reference values are undisclosed, while the reconstruction values represent attenuation coefficients.
These two issues preclude a meaningful voxel-wise comparison.
Consequently, we adopt a more reliable surface mesh-based evaluation.
Vessel surface meshes are extracted using the Marching Cubes algorithm~\citep{marchingcube}, with intensity thresholds of 0.04 for reference and 0.01 for reconstructions to account for the aforementioned intensity scale discrepancy.
Both thresholds are \R{determined via sensitivity analysis} to balance noise suppression and detail preservation, as validated in \cref{fig:thres_justify}.
Reconstruction meshes are aligned to reference using iterative closest point (ICP)~\citep{ICP}, markedly reducing surface distances (\cref{fig:why_mesh}(b)) and enabling reliable geometric evaluation independent of initial misalignment.
Chamfer distance (CD) and 95\% Hausdorff distance (HD95) are computed in millimeters as 3D metrics.
Meshes in \cref{fig:why_mesh}(b) are rendered with Open3D~\citep{Open3D} to illustrate the ICP alignment process and distance map comparison, while other 3D visualizations in our paper are generated with 3D Slicer~\citep{3D_slicer}.
For evaluating 2D image synthesis, we compare the synthesized DSA images with ground truth using Peak Signal-to-Noise Ratio (PSNR) and Structural Similarity Index Measure (SSIM)~\citep{SSIM}, with metrics for each case averaged over the test set.
The best performance in the quantitative tables is shown in bold.
Metrics (CD($\mathrm{mm}$)/HD95($\mathrm{mm}$), PSNR($\mathrm{dB}$)/SSIM) are shown at the top right of result panels.

\subsection{Model Output Analysis}
\label{sec:model output}

Given the complexity of our model outputs, especially the dynamic parts, we provide a comprehensive illustration through both 3D reconstructions and 2D renderings.
All \R{results} are from case \#1 with 40 training views.

\subsubsection{3D Vessel Reconstruction}
\label{sec:exp model out 3D}

\begin{figure}[t]
\centering
    \includegraphics[width=0.48\textwidth]{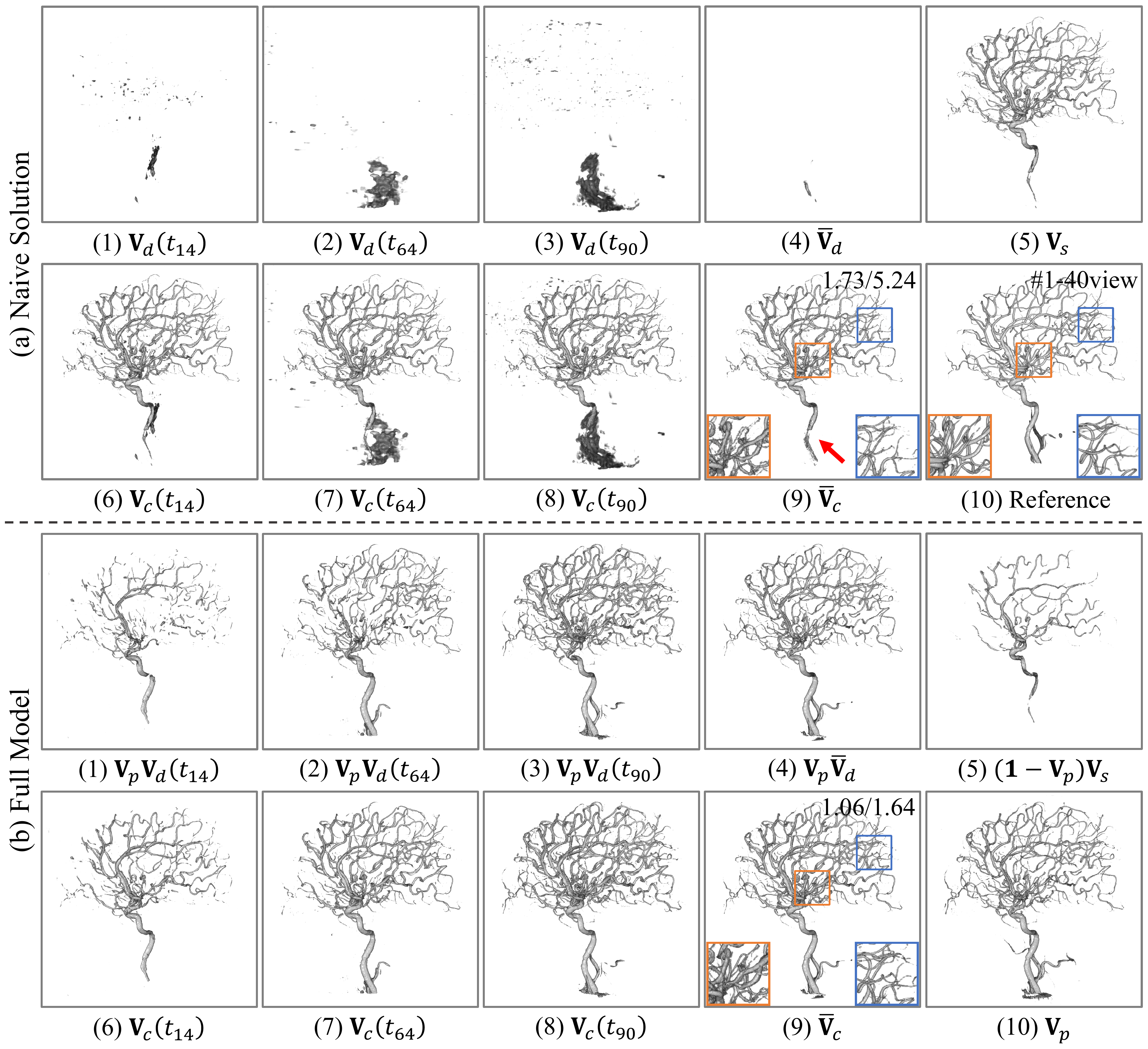}
    \caption{
    Decomposed 3D reconstruction.
    (a) Naive solution: volume of (1-3) dynamic attenuation $\mathbf{V}_d(t)$, (4) averaged dynamic attenuation $\overline{\mathbf{V}}_d$, (5) static attenuation $\mathbf{V}_s$, (6-8) contrast agent attenuation $\mathbf{V}_c(t)$, (9) reconstructed 3D vessel $\overline{\mathbf{V}}_c$, and (10) reference 3D vessel.
    (b) Full model: volume of (1-3) dynamic component $\mathbf{V}_p\mathbf{V}_d(t)$, (4) averaged dynamic component $\mathbf{V}_p\overline{\mathbf{V}}_d$, (5) static component $(\mathbf{1}-\mathbf{V}_p)\mathbf{V}_s$, (6-8) contrast agent attenuation $\mathbf{V}_c(t)$, (9) reconstructed 3D vessel $\overline{\mathbf{V}}_c$, and (10) vessel probability $\mathbf{V}_p$.
    Our model achieves a clear static-dynamic decomposition and high-quality vessel reconstruction, whereas the naive solution performs substantially worse.
    } 
    \label{fig:vol_compare}
    \vspace{-4mm}
\end{figure}

Our method derives various volumes as mentioned in \cref{sec:method vessel reconstruction}.
Vessel probability volume $\mathbf{V}_p$ (\cref{fig:vol_compare}(b,10)) captures meaningful vascular patterns.
It guides the decomposition of contrast flow into its static component $(\mathbf{1}-\mathbf{V}_p)\mathbf{V}_s$ (\cref{fig:vol_compare}(b,5)), and dynamic components $\mathbf{V}_p\mathbf{V}_d(t)$ (\cref{fig:vol_compare}(b,1-3)).
Their combination yields the time-varying contrast agent volumes $\mathbf{V}_c(t)$ (\cref{fig:vol_compare}(b,6-8)).
The final time-averaged 3D vessel reconstruction $\overline{\mathbf{V}}_c$ (\cref{fig:vol_compare}(b,9)) closely matches the reference volume (\cref{fig:vol_compare}(a,10)), achieving CD/HD95 of 1.06/1.64$\mathrm{mm}$.

To validate the efficacy of the proposed vessel probability field, we evaluated the naive solution lacking this component as discussed in \cref{sec:method naive solution}.
Without vessel probability guidance, the model fails to achieve a clean static-dynamic decomposition.
Consequently, the dynamic volumes $\mathbf{V}_d(t)$ and $\overline{\mathbf{V}}_d$ (\cref{fig:vol_compare}(a,1-4)) appear noisy without discernible vessel structures.
Both $\mathbf{V}_c(t)$ and $\overline{\mathbf{V}}_c$ (\cref{fig:vol_compare}(a,6-9)) are substantially degraded.
CD/HD95 deteriorate to 1.73/5.24$\mathrm{mm}$, respectively.
These results underscore the key role of vessel probability field in accurate decomposition and high-quality reconstruction.

\subsubsection{2D DSA Image Synthesis}
\label{sec:exp model out 2D}

\begin{figure*}[t]
\centering
    \includegraphics[width=\textwidth]{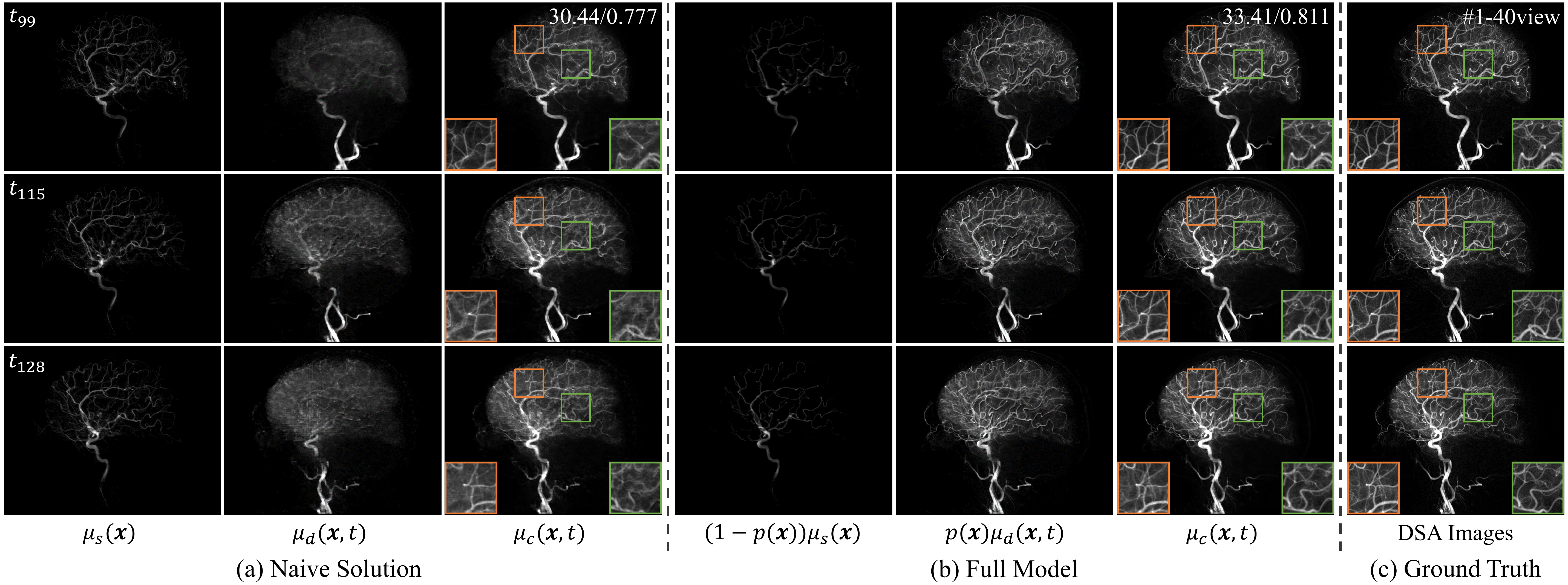}
    \caption{
Decomposed 2D rendering on test frames.
(a) Naive solution: rendering of static attenuation $\mu_s(\bm{x})$, dynamic attenuation $\mu_d(\bm{x},t)$, and synthesized DSA images $\mu_c(\bm{x},t)$. 
(b) Full model: rendering of static component $(1-p(\bm{x}))\mu_s(\bm{x})$, dynamic component $p(\bm{x})\mu_d(\bm{x},t)$, and synthesized DSA images $\mu_c(\bm{x},t)$.
(c) Ground truth DSA images.
Our full model significantly outperforms the naive solution, achieving clear static-dynamic decomposition and high-quality DSA image synthesis.
    } 
    \label{fig:proj_compare}
    \vspace{-4mm}
\end{figure*}

\begin{figure*}[th!]
\centering
    \includegraphics[width=1.0\textwidth]{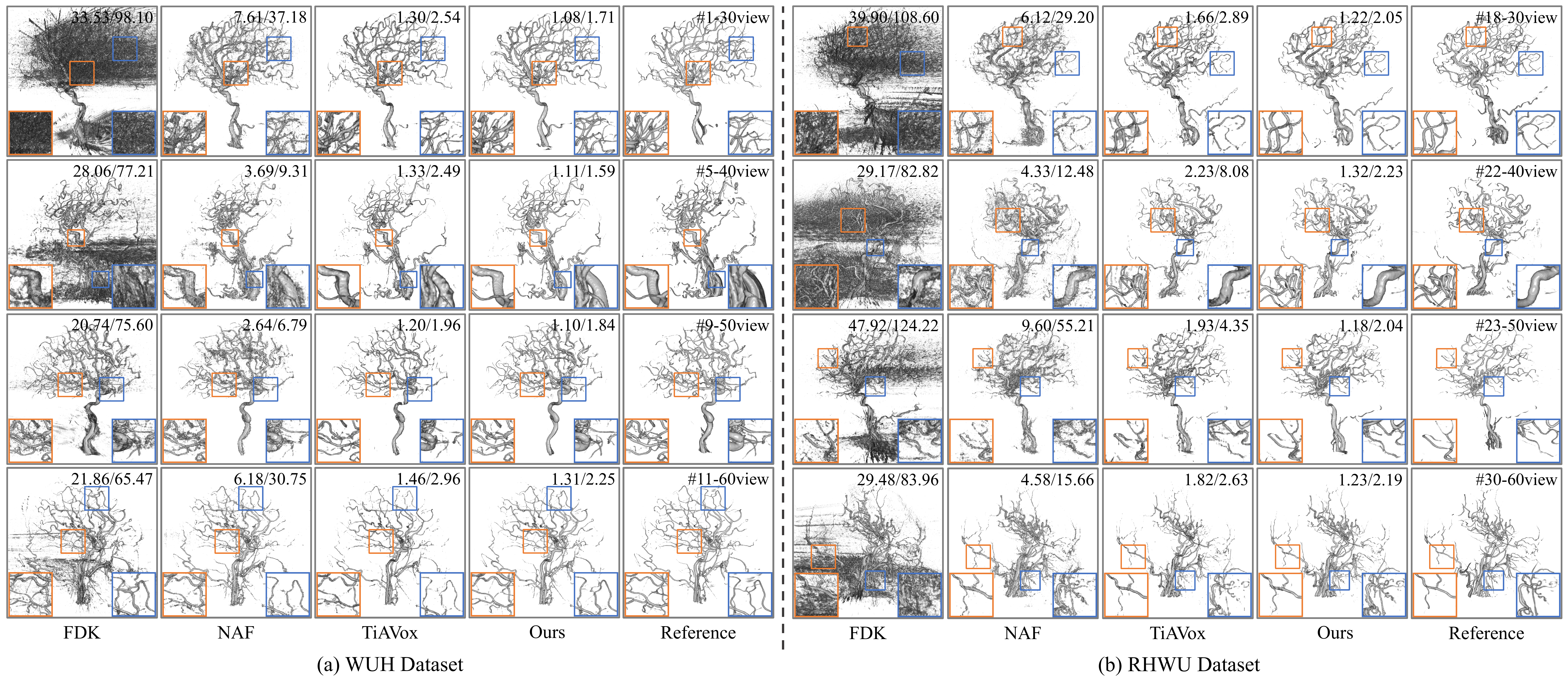}
    \caption{
    \R{
    3D vessel reconstruction comparison on different datasets.
    (a) Results on the WUH dataset. 
    (b) Results on the RHWU dataset.
    Our method consistently reconstructs clearer vessel structures with less noise and finer details across both datasets compared to competing approaches.
    }
    } 
    \label{fig:results_vol_compare}
    \vspace{-4mm}
\end{figure*}

\begin{figure*}[th!]
\centering
    \includegraphics[width=1.0\textwidth]{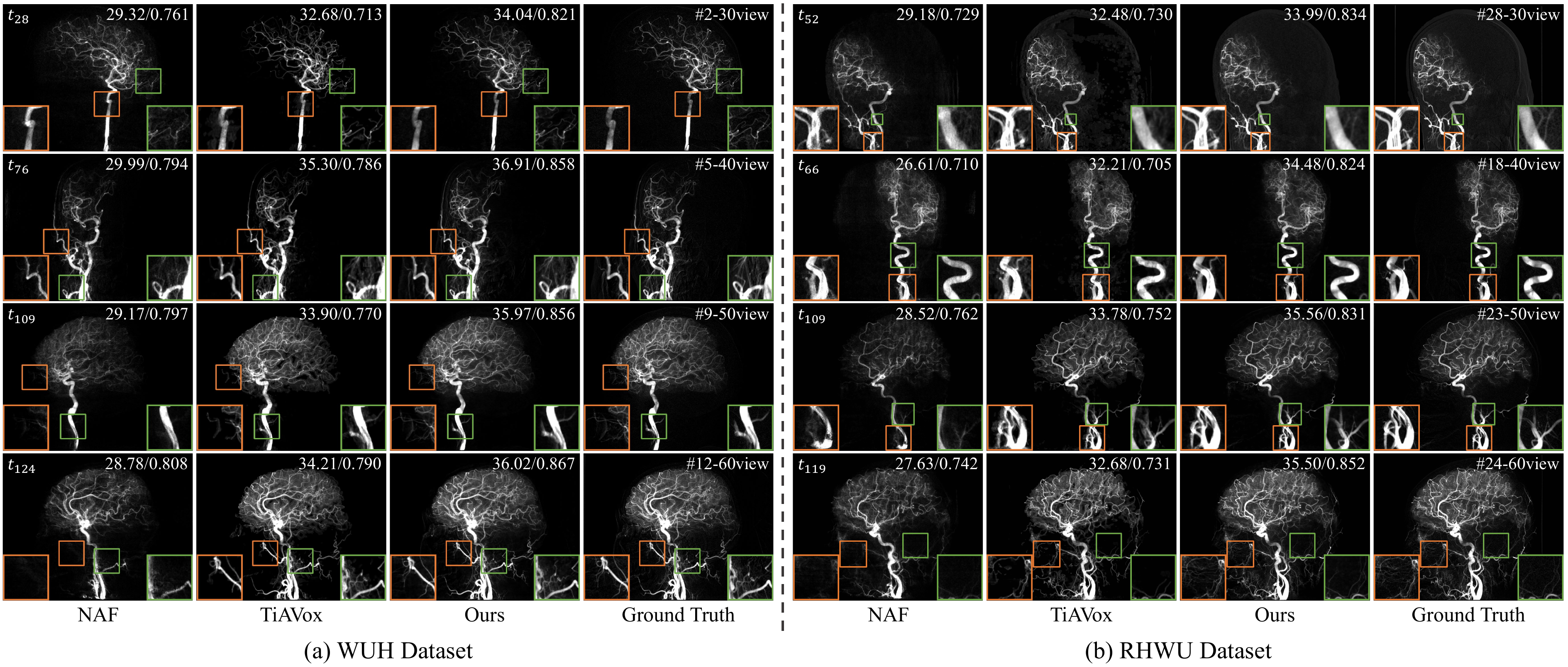}
    \caption{
    \R{
    2D DSA synthesis comparison on different datasets.
    (a) Results on the WUH dataset. 
    (b) Results on the RHWU dataset.
    Cases \#24 and \#28 ($960\times960$) are horizontally zero-padded to match other cases' projection resolution ($1240\times960$) for layout consistency.
   Our method achieves high-quality DSA synthesis, exhibiting reduced artifacts and finer details compared to competing approaches.
    }
    } 
    \label{fig:results_proj_compare}
    \vspace{-4mm}
\end{figure*}

Our model synthesizes 2D DSA images given desired viewpoints and timestamps via volume rendering with $\mu_c(\bm{x}, t)$ (\cref{eq:modelrendering}), and can also render the decomposed static $(1-p(\bm{x}))\mu_s(\bm{x})$ and dynamic $p(\bm{x})\mu_d(\bm{x},t)$ components.
The rendered projections (\cref{fig:proj_compare}(b)) demonstrate a clean static-dynamic separation and exhibit high fidelity DSA image synthesis compared to ground truth (\cref{fig:proj_compare}(c)), achieving PSNR/SSIM of 33.41$\mathrm{dB}$/0.811.

In contrast, rendering $\mu_s(\bm{x})$, $\mu_d(\bm{x},t)$, and $\mu_c(\bm{x}, t)$ with the naive solution yields significantly different results as shown in \cref{fig:proj_compare}(a).
The static-dynamic decomposition appears ambiguous, resulting in inferior synthesized DSA images with blurring and loss of details.
PSNR/SSIM degrade to 30.44$\mathrm{dB}$/0.777, respectively.
This further demonstrates the effectiveness of our proposed vessel probability field.

\subsection{Experimental Results}

\subsubsection{Quantitative Evaluation}
\label{sec:exp quantitative}

\begin{table}[]
\caption{\R{Quantitative comparison on the WUH dataset. Our method consistently outperforms competing approaches, achieving superior metrics in both 3D vessel reconstruction (CD, HD95) and 2D DSA image synthesis (PSNR, SSIM) across all sparse-view settings.}}
\centering
\fontsize{6.5}{6.5}\selectfont
\setlength{\tabcolsep}{6pt}
\renewcommand\arraystretch{1.2}
\begin{tabular}{cccccc}
\toprule
Views               & Method & CD ($\mathrm{mm}$) $\downarrow$ & HD95 ($\mathrm{mm}$) $\downarrow$ & PSNR ($\mathrm{dB}$) $\uparrow$ & SSIM $\uparrow$          \\ \midrule
\multirow{4}{*}{30} & FDK    & 30.26$\pm$4.95                  & 87.66$\pm$11.02                 & -                               & -                        \\
                    & NAF    & 7.40$\pm$2.65                   & 31.85$\pm$18.90                 & 28.37$\pm$0.98                  & 0.741$\pm$0.043          \\
                    & TiAVox & 1.99$\pm$0.67                   & 5.00$\pm$3.34                 & 32.24$\pm$1.27                  & 0.729$\pm$0.045          \\
                    & Ours   & \textbf{1.30$\pm$0.15}          & \textbf{2.55$\pm$0.90}          & \textbf{33.71$\pm$1.38}         & \textbf{0.825$\pm$0.031} \\ \midrule
\multirow{4}{*}{40} & FDK    & 27.67$\pm$5.04                  & 83.63$\pm$11.29                 & -                               & -                        \\
                    & NAF    & 6.62$\pm$2.50                   & 30.60$\pm$21.18                 & 28.53$\pm$0.97                  & 0.750$\pm$0.042          \\
                    & TiAVox & 1.72$\pm$0.49                   & 3.70$\pm$1.72                 & 32.70$\pm$1.33                  & 0.735$\pm$0.044          \\
                    & Ours   & \textbf{1.28$\pm$0.15}          & \textbf{2.33$\pm$0.74}          & \textbf{34.43$\pm$1.35}         & \textbf{0.833$\pm$0.029} \\ \midrule
\multirow{4}{*}{50} & FDK    & 24.97$\pm$5.62                  & 79.25$\pm$13.37                 & -                               & -                        \\
                    & NAF    & 5.47$\pm$2.58                   & 22.43$\pm$19.76                 & 28.73$\pm$1.02                  & 0.757$\pm$0.041          \\
                    & TiAVox & 1.60$\pm$0.41                   & 3.29$\pm$1.26                 & 32.96$\pm$1.34                  & 0.738$\pm$0.044          \\
                    & Ours   & \textbf{1.26$\pm$0.18}          & \textbf{2.40$\pm$1.11}          & \textbf{34.95$\pm$1.40}         & \textbf{0.839$\pm$0.028} \\ \midrule
\multirow{4}{*}{60} & FDK    & 22.30$\pm$5.87                  & 75.57$\pm$14.97                 & -                               & -                        \\
                    & NAF    & 5.27$\pm$2.59                   & 27.46$\pm$25.62                 & 28.80$\pm$1.00                  & 0.760$\pm$0.038          \\
                    & TiAVox & 1.54$\pm$0.38                   & 3.06$\pm$1.21                 & 33.07$\pm$1.29                  & 0.739$\pm$0.042          \\
                    & Ours   & \textbf{1.24$\pm$0.13}          & \textbf{2.25$\pm$0.71}          & \textbf{35.28$\pm$1.37}         & \textbf{0.842$\pm$0.025} \\ \bottomrule
\end{tabular}
\label{table:Quantitative Results WUH}
\vspace{-2mm}
\end{table}

\begin{table}[t]
% \color{red}
\caption{\R{Quantitative comparison on the RHWU dataset. Consistent with the WUH results, our method maintains the best performance across all metrics and view settings, demonstrating strong generalizability.}}
\centering
\fontsize{6.5}{6.5}\selectfont
\setlength{\tabcolsep}{6pt}
\renewcommand\arraystretch{1.2}
\begin{tabular}{cccccc}
\toprule
Views               & Method & CD ($\mathrm{mm}$) $\downarrow$ & HD95 ($\mathrm{mm}$) $\downarrow$ & PSNR ($\mathrm{dB}$) $\uparrow$ & SSIM $\uparrow$          \\ \midrule
\multirow{4}{*}{30} & FDK    & 30.59$\pm$9.80                  & 89.90$\pm$21.31                   & -                               & -                        \\
                    & NAF    & 6.91$\pm$2.45                   & 36.39$\pm$18.49                   & 28.49$\pm$1.19                  & 0.749$\pm$0.043          \\
                    & TiAVox & 1.86$\pm$0.26                   & 3.87$\pm$1.70                     & 32.72$\pm$1.18                  & 0.760$\pm$0.038          \\
                    & Ours   & \textbf{1.23$\pm$0.14}          & \textbf{2.34$\pm$0.93}            & \textbf{34.06$\pm$1.19}         & \textbf{0.841$\pm$0.024} \\ \midrule
\multirow{4}{*}{40} & FDK    & 27.66$\pm$10.86                 & 85.96$\pm$23.15                   & -                               & -                        \\
                    & NAF    & 5.74$\pm$3.01                   & 26.47$\pm$22.96                   & 28.65$\pm$1.15                  & 0.757$\pm$0.041          \\
                    & TiAVox & 1.69$\pm$0.24                   & 3.51$\pm$1.52                     & 33.25$\pm$1.18                  & 0.765$\pm$0.039          \\
                    & Ours   & \textbf{1.22$\pm$0.14}          & \textbf{2.27$\pm$0.90}            & \textbf{34.91$\pm$1.15}         & \textbf{0.850$\pm$0.023} \\ \midrule
\multirow{4}{*}{50} & FDK    & 24.18$\pm$12.23                 & 80.37$\pm$25.77                   & -                               & -                        \\
                    & NAF    & 4.64$\pm$3.03                   & 18.84$\pm$19.41                   & 28.82$\pm$1.14                  & 0.764$\pm$0.039          \\
                    & TiAVox & 1.60$\pm$0.21                   & 3.30$\pm$1.37                     & 33.54$\pm$1.14                  & 0.770$\pm$0.037          \\
                    & Ours   & \textbf{1.29$\pm$0.45}          & \textbf{2.66$\pm$1.94}            & \textbf{35.01$\pm$1.80}         & \textbf{0.851$\pm$0.026} \\ \midrule
\multirow{4}{*}{60} & FDK    & 20.65$\pm$13.02                 & 73.37$\pm$30.69                   & -                               & -                        \\
                    & NAF    & 4.78$\pm$3.77                   & 20.29$\pm$24.77                   & 28.90$\pm$1.12                  & 0.767$\pm$0.039          \\
                    & TiAVox & 1.55$\pm$0.18                   & 3.28$\pm$1.44                     & 33.71$\pm$1.12                  & 0.772$\pm$0.036          \\
                    & Ours   & \textbf{1.18$\pm$0.15}          & \textbf{2.20$\pm$0.95}            & \textbf{35.84$\pm$1.24}         & \textbf{0.859$\pm$0.022} \\ \bottomrule
\end{tabular}
\label{table:Quantitative Results RHWU}
\vspace{-4mm}
\end{table}

% \cref{table:Quantitative Results} provides quantitative comparison between different methods.
% Our method demonstrates promising performance, particularly excelling in 3D vessel reconstruction metrics.

\R{\cref{table:Quantitative Results WUH} and \cref{table:Quantitative Results RHWU} report the quantitative results on the WUH and RHWU datasets, respectively.
Across both clinical centers, our method consistently achieves the best performance across all objective evaluated metrics.}
Both FDK and NAF are static methods that do not model DSA dynamics, resulting in poor reconstruction and rendering performance. 
TiAVox achieves notable metric improvements over FDK and NAF, but its performance is still constrained by inefficient temporal modeling via 4D voxel grids.
Our method clearly surpasses these competitors, demonstrating our effectiveness on both 3D vessel reconstruction and 2D DSA image synthesis quality.
\R{Moreover, the consistent superiority on both datasets demonstrates the robustness and generalizability of the proposed framework.}

\subsubsection{Qualitative Evaluation}
\label{sec:exp qualitative}

\cref{fig:results_vol_compare} presents qualitative comparison on 3D vessel reconstruction.
FDK suffers from severe noise and streak artifacts due to its static nature and limited data.
NAF \R{improves structural recovery over FDK via} NeRF-based optimization but also ignores temporal dynamics, leading to detail loss and noise.
TiAVox employs learnable 4D voxel grids to model dynamic DSA sequences and achieves decent results.
Yet its reconstructions still exhibit noise and detail loss due to inefficient temporal modeling.
In contrast, our approach delivers superior reconstruction quality, with smooth surfaces, low noise, and detailed vessels that closely match the reference.

Qualitative comparison of 2D DSA image synthesis on test frames is shown in \cref{fig:results_proj_compare}.
NAF fails to capture vessel details and introduces noise due to its inability to model DSA dynamics.
TiAVox generates \R{visually plausible renderings but still} introduces artifacts and loses details as marked by colored boxes, yielding inferior quality compared to our approach.
Our method effectively models the dynamic nature of DSA sequence, achieving superior quality in DSA image synthesis.

\subsubsection{\R{Clinical Expert Evaluation}}
\label{sec:expert_score}

\begin{table}[]
% \color{red}
\caption{\R{Expert scoring results (30-view). VDV: Vessel detail visibility; NAS: Noise and artifact suppression; DC: Diagnostic confidence. Our method consistently outperforms competitors in all subjective metrics.}}
\centering
\fontsize{6.5}{6.5}\selectfont
\setlength{\tabcolsep}{5pt}
\renewcommand\arraystretch{1.2}
\begin{tabular}{cccc|ccc}
\toprule
Dataset   & \multicolumn{3}{c|}{WUH}                         & \multicolumn{3}{c}{RHWU}   \\ \midrule
Method    & VDV $\uparrow$ & NAS $\uparrow$ & DC $\uparrow$ & VDV $\uparrow$    & NAS $\uparrow$   & DC $\uparrow$   \\ \midrule
FDK       & 1.13$\pm$0.34          & 1.07$\pm$0.25              & 1.05$\pm$0.22              & 1.13$\pm$0.38             & 1.05$\pm$0.22             & 1.07$\pm$0.25                \\
NAF       & 2.45$\pm$0.79          & 2.11$\pm$0.62          & 2.29$\pm$0.69          & 2.13$\pm$0.68             & 1.89$\pm$0.53            & 1.93$\pm$0.64            \\
TiAVox    & 3.19$\pm$0.76          & 3.05$\pm$0.75          & 3.12$\pm$0.80          & 3.11$\pm$0.70             & 2.99$\pm$0.66            & 3.00$\pm$0.65            \\
Ours      & \textbf{3.93$\pm$0.74}          & \textbf{3.95$\pm$0.76}          & \textbf{3.88$\pm$0.67}          & \textbf{3.92$\pm$0.58}             & \textbf{3.81$\pm$0.72}            & \textbf{3.93$\pm$0.60}            \\ \bottomrule
\end{tabular}
\label{table:exper_score}
\vspace{-6mm}
\end{table}

\R{Since absolute ground truth is unavailable in clinical settings and scanner-provided FDK reconstructions exhibit imperfections as shown in \cref{fig:imperfect_reference}, we conducted a blinded medical expert study to further evaluate the real-world clinical utility of our method.
Given the substantial evaluation effort, we specifically focused on the most challenging 30-view scenario.
Five neuro-interventional radiologists, each with at least three years of experience, unbiasedly scored the 3D vessel reconstruction quality on a five-point Likert scale based on three critical criteria: vessel detail visibility (VDV), noise and artifact suppression (NAS), and diagnostic confidence (DC).
As summarized in \cref{table:exper_score}, our method consistently achieved the highest ratings across all subjective metrics, significantly outperforming competing approaches. 
These findings confirm that our framework effectively preserves fine vascular details while suppressing artifacts, offering superior reliability for clinical decision-making. 
This subjective trend aligns well with the objective quantitative metrics and qualitative visual results presented above.}

% \subsubsection{Runtime Performance}

% In terms of per-case runtime, FDK completes reconstruction in seconds, NAF \textasciitilde30 minutes, TiAVox \textasciitilde6 minutes, and our method takes \textasciitilde2 hours.
% Our slow training arises from two factors of the model design.
% First, volumetric rendering is computationally intensive due to massive sample points evaluated per casting ray.
% Second, our model architecture is highly expressive, consisting of three neural fields to model the entire space.
% Thus, the per-sample-point network computation is relatively slow.
% We discuss the avenues for runtime acceleration and the potential for real-time inference in \cref{sec:discussion}.

\subsubsection{\R{Efficiency Analysis}}
\label{sec:efficiency}

\begin{table}[]
% \color{red}
\centering
\fontsize{8}{8}\selectfont
\setlength{\tabcolsep}{6pt}
\renewcommand\arraystretch{1.2}
\caption{\R{Computational efficiency analysis (40-view).
Runtime statistics are computed over all 30 cases.
Our method maintains a moderate model size and memory usage suitable for clinical workstations. M: millions of parameters.}}
\begin{tabular}{cccc}
\toprule
       & Runtime & Model Size (M) & Memory Usage (GB) \\ \midrule
FDK    & 1.24s$\pm$0.22s    & -              & -                 \\
NAF    & 33m12s$\pm$56s   & 35.35          & 11.21              \\
TiAVox & 6m48s$\pm$23s    & 228.04         & 12.64             \\
Ours   & 2h2m$\pm$6m18s     & 71.53          & 11.23              \\ \bottomrule
\end{tabular}
\label{table:computation}
\vspace{-4mm}
\end{table}

\R{To evaluate computational efficiency, we report runtime, model parameter size, and GPU memory usage for the 40-view setting in \cref{table:computation}.
Runtime statistics are computed over all 30 cases.
Our method maintains a moderate model size and memory usage, making it deployable on clinical workstations.}

\R{We acknowledge that the per-case runtime of our method ($\sim$2 hours) is relatively long compared with other approaches.
The runtime overhead mainly comes from computationally intensive volumetric rendering and the use of three neural fields.
Our framework prioritizes high-quality sparse-view reconstruction. 
This makes it suitable for non-acute workflows including preoperative planning and hemodynamic analysis, rather than acute interventions such as stroke thrombectomy.}
We discuss avenues for runtime acceleration and the potential for real-time inference in \cref{sec:discussion}.

\subsection{Ablation Study}

We perform thorough ablation studies on key model components, regularization \R{weight}, perturbation size, and hash encoder hyperparameters.
All ablations are conducted with 40 input views \R{on the WUH dataset}.
In \cref{table:ab_lreg,table:ab_perturbation_size,table:ab_hash_params}, \textsuperscript{$\dag$} marks the default hyperparameter setting used in our model.

% , and further discuss an alternative solution.

\subsubsection{Model Components}
\label{sec:ab components}

\begin{table}[t]
\caption{
\R{Quantitative results of ablation study on model components.
$\{\mathcal{S},\mathcal{D}\}$: Naive solution without vessel probability field $\mathcal{P}$;
PG: Coarse-to-fine progressive training; 
TP: Temporal perturbed rendering loss.}}
\centering
\fontsize{6.5}{6.5}\selectfont
\setlength{\tabcolsep}{4pt}
\renewcommand\arraystretch{1.2}
\begin{tabular}{cccccccc}
\toprule
$\{\mathcal{S},\mathcal{D}\}$ & $\mathcal{P}$          & PG           & TP           & CD ($\mathrm{mm}$) $\downarrow$ & HD95 ($\mathrm{mm}$) $\downarrow$ & PSNR ($\mathrm{dB}$) $\uparrow$ & SSIM $\uparrow$ \\ \midrule
$\checkmark$  &              &              &              & 3.04$\pm$2.53            & 10.94$\pm$9.00           & 30.61$\pm$1.20           & 0.786$\pm$0.034           \\
$\checkmark$  & $\checkmark$ &              &              & 1.37$\pm$0.18            & 3.19$\pm$0.92            & 34.08$\pm$1.42           & 0.823$\pm$0.032           \\
$\checkmark$  & $\checkmark$ & $\checkmark$ &              & 1.31$\pm$0.17            & 2.61$\pm$0.91            & 34.26$\pm$1.46           & 0.828$\pm$0.030           \\
$\checkmark$  & $\checkmark$ & $\checkmark$ & $\checkmark$ & \textbf{1.28$\pm$0.15}            & \textbf{2.33$\pm$0.74}            & \textbf{34.43$\pm$1.35}           & \textbf{0.833$\pm$0.029}           \\ \bottomrule
\end{tabular}
\label{table:ab_model_components}
\vspace{-2mm}
\end{table}

\begin{figure}[t]
\centering
    \includegraphics[width=0.48\textwidth]{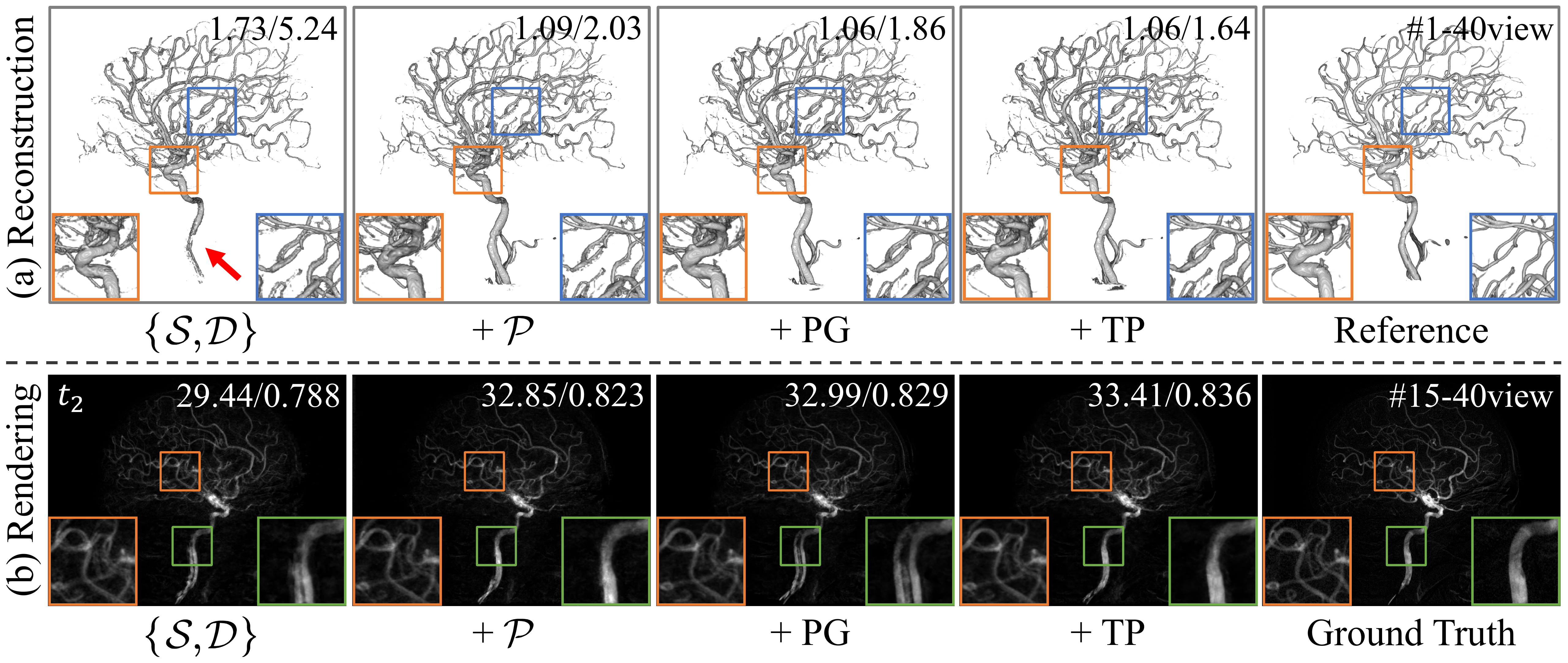}
    \caption{
    \R{
    Qualitative ablation results on model components.
    $\{\mathcal{S},\mathcal{D}\}$: Naive solution; 
    $\mathcal{P}$: Vessel probability field;
    PG: Progressive training;
    TP: Temporal perturbed loss.
    (a) 3D vessel reconstruction.
    (b) 2D DSA synthesis on test frame. 
    } 
    } 
    \label{fig:ab_model_components}
    \vspace{-4mm}
\end{figure}

We study the effectiveness of vessel probability field and training strategies. 
\cref{table:ab_model_components} reports the quantitative results.
Here, $\{\mathcal{S},\mathcal{D}\}$ denotes the naive solution, $\mathcal{P}$ represents vessel probability field, while PG and TP stand for coarse-to-fine progressive training and temporal perturbed rendering loss, respectively.
A checkmark $\checkmark$ indicates the inclusion of the specific component.
The naive solution performs the worst in both 3D and 2D quality.
Introducing the vessel probability field leads to notable metric improvements, highlighting its importance, as discussed in \cref{sec:model output}.
The incorporation of PG and TP further boosts our results, confirming the effectiveness of these two training strategies.

\cref{fig:ab_model_components}(a) displays 3D vessel reconstruction results.
The inclusion of the vessel probability field leads to more complete vascular structures compared to the naive solution, but also introduces surface noise, as marked by the orange box.
This is attributed to the increased model complexity brought by the additional parameters of vessel probability field.
Given a fixed number of training views, the model becomes more underdetermined, which reduces solution stability and increases noise.
Progressive training effectively suppresses artifacts, yielding smoother surfaces and improved geometric accuracy.
It promotes stable convergence by guiding the model to first capture low-frequency structures and then refine high-frequency details, thus reducing noise.

\cref{fig:ab_model_components}(b) shows 2D DSA synthesis results on a test frame.
While progressive training improves stability, its low-frequency prior optimization compromises the modeling of \R{rapid contrast flow in large arteries}.
This may cause overfitting on sparse training frames and introduce temporally inconsistent floating artifacts, as highlighted in the green box.
Notably, the inclusion of temporal perturbation significantly reduces these discontinuous artifacts by encouraging temporal consistency between neighboring timestamps.
For a detailed analysis of how temporal perturbation affects model performance, please refer to \cref{sec:ab perturbation size}.

\subsubsection{Regularization \R{Weight}}

\begin{table}[t]
\caption{Quantitative results of ablation study on regularization \R{weight $\lambda_{\text{reg}}$.}}
\centering
\fontsize{6.5}{6.5}\selectfont
\setlength{\tabcolsep}{10pt}
\renewcommand\arraystretch{1.2}
\begin{tabular}{ccccc}
\toprule
$\lambda_{\text{reg}}$ & CD ($\mathrm{mm}$) $\downarrow$ & HD95 ($\mathrm{mm}$) $\downarrow$ & PSNR ($\mathrm{dB}$) $\uparrow$ & SSIM $\uparrow$      \\ \midrule
0                    & 1.46$\pm$0.23                   & 3.52$\pm$1.54                   & 34.03$\pm$1.47                  & 0.826$\pm$0.031      \\
0.001                  & 1.30$\pm$0.19            & 2.53$\pm$1.22            & 34.38$\pm$1.37            & 0.831$\pm$0.029 \\
0.01\textsuperscript{$\dag$}                   & \textbf{1.28$\pm$0.15}                   & \textbf{2.33$\pm$0.74}                   & \textbf{34.43$\pm$1.35}                  & \textbf{0.833$\pm$0.029}      \\
0.1                    & 1.31$\pm$0.19            & 2.55$\pm$1.17            & 34.30$\pm$1.36           & 0.830$\pm$0.029\\ \bottomrule
\end{tabular}
\label{table:ab_lreg}
\vspace{-2mm}
\end{table}

% \begin{figure}[th!]
% \centering
%     \includegraphics[width=0.48\textwidth]{Img/ab_Lreg.pdf}
%     \caption{
%     Qualitative results of ablation study on regularization term. 
%     (a) Decomposed 3D reconstruction: volume of vessel probability $\mathbf{V}_p$, static component $(\mathbf{1}-\mathbf{V}_p)\mathbf{V}_s$, averaged dynamic component $\mathbf{V}_p\overline{\mathbf{V}}_d$, and reconstructed 3D vessel $\overline{\mathbf{V}}_c$. 
%     (b) Decomposed 2D rendering on a test frame: rendering of static component $(1-p(\bm{x}))\mu_s(\bm{x})$, dynamic component $p(\bm{x})\mu_d(\bm{x},t)$, and synthesized DSA $\mu_c(\bm{x},t)$.
%     } 
%     \label{fig:ab_Lreg}
%     \vspace{-4mm}
% \end{figure}

\begin{figure}[t]
\centering
    \includegraphics[width=0.48\textwidth]{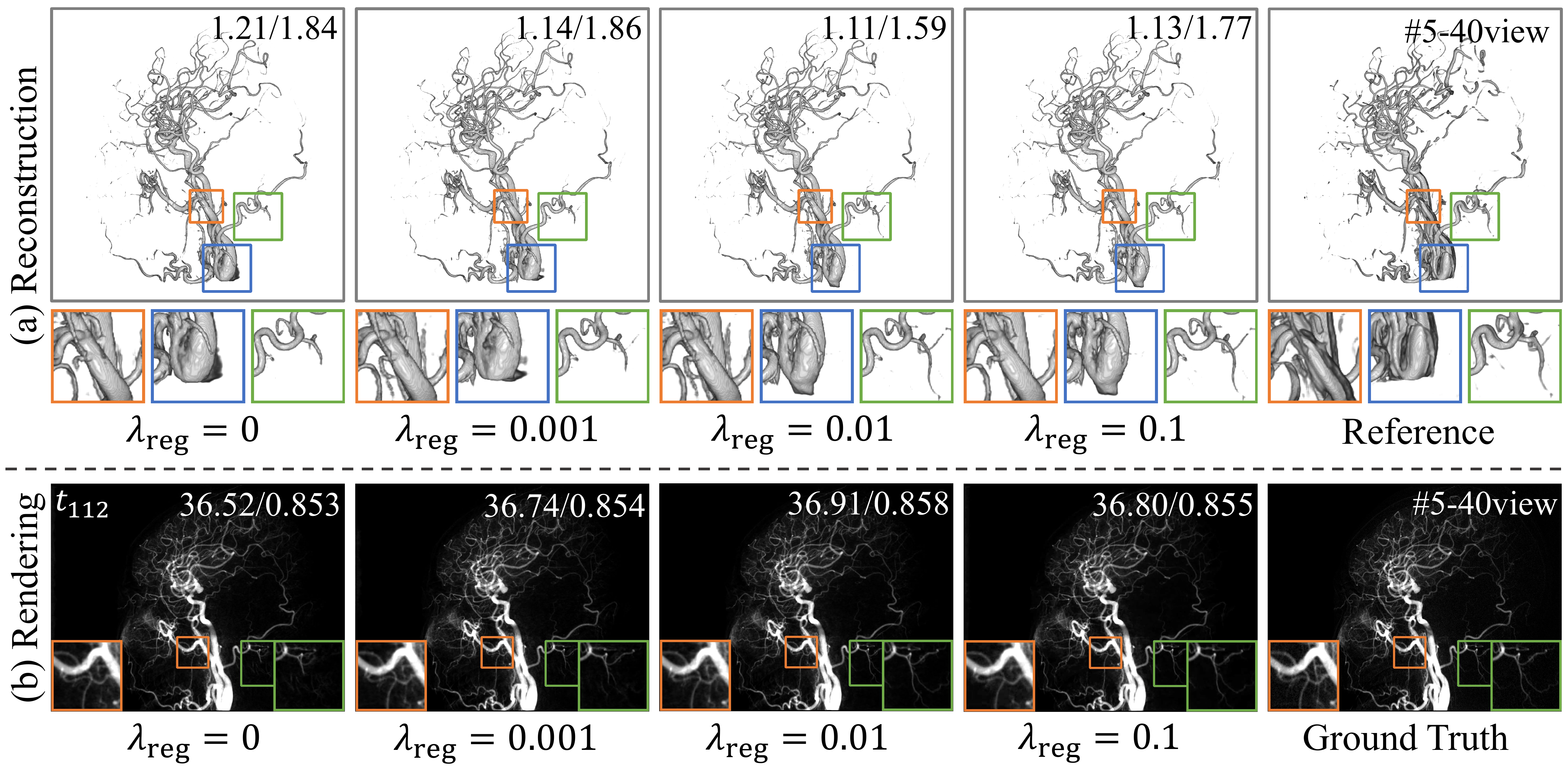}
    \caption{
    \R{
    Qualitative results of ablation study on regularization weight $\lambda_{\text{reg}}$.
    (a) 3D vessel reconstruction.
    (b) 2D DSA image synthesis on test frame.
    }
    } 
\label{fig:ab_lambdareg}
\vspace{-4mm}
\end{figure}

% We investigate the effect of the regularization term.
% As shown in \cref{table:ab_lreg}, removing it degrades performance across all evaluation metrics.
% \cref{fig:ab_Lreg}(a) presents 3D reconstruction results with decomposed components. 
% Without regularization, the vessel probability volume $\mathbf{V}_p$ becomes significantly noisy, impairing the static-dynamic decomposition and leading to noticeable surface noise in the reconstructed vessel as marked by the orange box.
% \cref{fig:ab_Lreg}(b) shows 2D renderings on a test frame, where the absence of regularization similarly hinders accurate decomposition and degrades synthesized DSA quality.
% We further verify the selection of regularization weight $\lambda_{\text{reg}}$ in \cref{table:ab_lreg} and \cref{fig:ab_lambdareg}.

\R{We investigate the impact of the regularization weight $\lambda_{\text{reg}}$ to justify its selection.
As shown in \cref{table:ab_lreg}, our chosen value $\lambda_{\text{reg}}=0.01$ yields the best performance across all four metrics, validating its effectiveness.
Visual comparisons in \cref{fig:ab_lambdareg} further support this selection.
When the regularization is absent or too weak ($\lambda_{\text{reg}} = 0$), the model fails to adequately constrain density responses in non-vascular regions.
Consequently, pronounced surface noise appears in the 3D reconstruction (blue box), while small vessel structures are partially lost in both 3D and 2D results (orange and green boxes).
In contrast, an excessively strong regularization ($\lambda_{\text{reg}} = 0.1$) tends to suppress the reconstruction of small vessels, as highlighted in the orange boxes.
Overall, our chosen setting $\lambda_{\text{reg}} = 0.01$ achieves the best performance, effectively reducing surface noise and improving small-vessel visibility.}

\subsubsection{Perturbation size}
\label{sec:ab perturbation size}

\begin{table}[t]
\caption{Quantitative results of ablation study on perturbation size \R{$k$}.
\R{DLE: Dynamics learning error.}}
\centering
\fontsize{6.5}{6.5}\selectfont
\setlength{\tabcolsep}{5.5pt}
\renewcommand\arraystretch{1.2}
\begin{tabular}{cccccc}
\toprule
$k$ & CD ($\mathrm{mm}$) $\downarrow$ & HD95 ($\mathrm{mm}$) $\downarrow$ & PSNR ($\mathrm{dB}$) $\uparrow$ & SSIM $\uparrow$ & DLE ($\times 10^{-4}$) $\downarrow$ \\ \midrule
0   & 1.31$\pm$0.17                                & 2.61$\pm$0.91                                & 34.26$\pm$1.46                                & 0.828$\pm$0.030                &  4.42$\pm$1.13                \\
0.1 & 1.29$\pm$0.15                               & 2.33$\pm$0.78                                & 34.27$\pm$1.43                                & 0.829$\pm$0.029                & 4.25$\pm$1.11                 \\
0.3 & 1.31$\pm$0.16                                & 2.34$\pm$0.74                                & 34.32$\pm$1.40                                & 0.830$\pm$0.029                & 4.26$\pm$1.11                 \\
0.5 & 1.31$\pm$0.16                                & 2.36$\pm$0.77                                & 34.40$\pm$1.39                                & 0.831$\pm$0.029                & 4.22$\pm$1.11                 \\
1\textsuperscript{$\dag$}   & \textbf{1.28$\pm$0.15}                                & \textbf{2.33$\pm$0.74}                                & \textbf{34.43$\pm$1.35}                                & \textbf{0.833$\pm$0.029}                & \textbf{4.13$\pm$1.09}                 \\
2   & 1.30$\pm$0.15                              & 2.35$\pm$0.74                                & 34.33$\pm$1.36                                & \textbf{0.833$\pm$0.029}                & 4.15$\pm$1.04                 \\
3   & 1.30$\pm$0.15                                & 2.37$\pm$0.78                                & 34.17$\pm$1.36                                & 0.832$\pm$0.030                & 4.22$\pm$1.09                 \\ \bottomrule
\end{tabular}
\label{table:ab_perturbation_size}
\vspace{-2mm}
\end{table}

\begin{figure}[t]
\centering
    \includegraphics[width=0.48\textwidth]{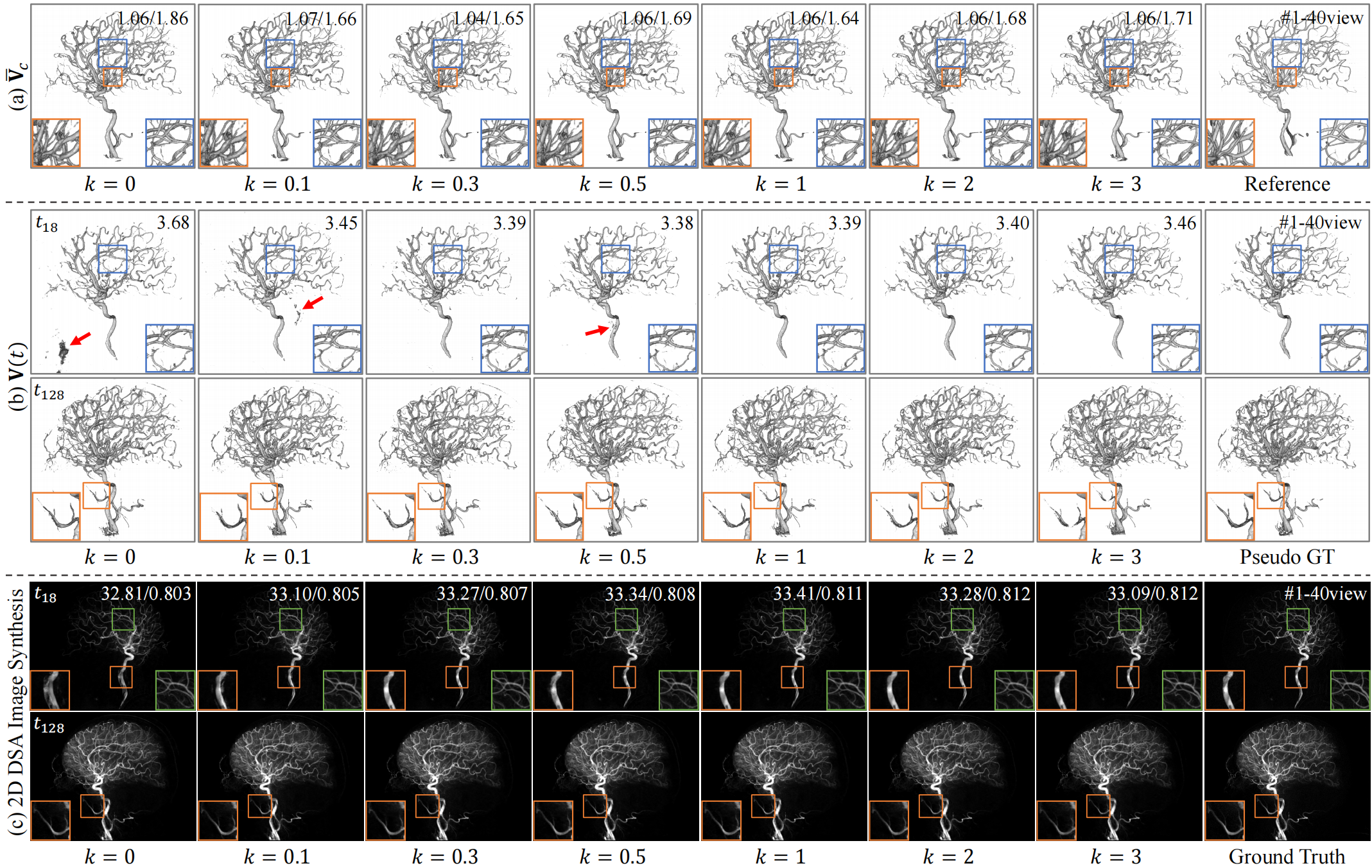}
    \caption{
    Qualitative results of ablation study on perturbation size $k$.
    (a) 3D vessel reconstruction $\overline{\mathbf{V}}_c$. 
    (b) Contrast volume $\mathbf{V}(t)$ on test frames with DLE ($\times10^{-4}$) top-right. \R{DLE: Dynamics learning error.}
    (c) 2D DSA image synthesis on test frames.
    }
    \label{fig:ab_perturbation_size}
    \vspace{-4mm}
\end{figure}

As shown in \cref{sec:ab components}, our temporal perturbation strategy enhances temporal consistency and reduces overfitting artifacts.
However, excessive perturbation might also over-smooth genuine blood flow dynamics, particularly in fast-flow vessels.
Essentially, temporal perturbation acts as a temporal low-pass filter for both signal and noise, where perturbation size $k$ controls smoothing strength.
To explore this trade-off, we thoroughly ablate parameter $k$.

We introduce Dynamics Learning Error (DLE) metric to quantify learned dynamics error that may be caused by perturbation mechanism.
We first establish a pseudo ground truth by training a reference model on full view set (133 views) with the temporal perturbation disabled, yielding reference contrast volumes $\{\mathbf{V}^{\text{pse}}_c(t_i)\in \mathbb{R}^{W \times H \times D}\}_{i=1}^{T}$.
As temporal perturbation alleviates sparse-view overfitting, disabling it with full data allows this model to capture genuine data dynamics for theoretically optimal reconstructions.
This compensates for the lack of time-resolved reference volumes in scanner data.
DLE is then defined as the time-averaged Mean Absolute Error (MAE) between our sparse-view model's volumes $\{\mathbf{V}_c(t_i)\}_{i=1}^{T}$ and this pseudo ground truth, restricted to binary mask $\mathbf{M}(t_i) \in \{0,1\}^{W \times H \times D}$ for meaningful computations:
\begin{equation}
\mathbf{M}(t_i) = (\mathbf{V}_c(t_i) > \epsilon) \lor (\mathbf{V}^{\text{pse}}_c(t_i) > \epsilon),
\label{eq:DLE_mask}
\end{equation}
\begin{equation} 
\text{DLE} = \frac{1}{T}\sum_{i=1}^{T} \mathbb{E}_{\mathbf{M}(t_i)}\left[ \left| \mathbf{V}_c(t_i) - \mathbf{V}^{\text{pse}}_c(t_i) \right| \right],
\label{eq:DLE} 
\end{equation}
\R{where} $\epsilon=10^{-6}$ filters zero voxels, $\lor$ denotes voxel-wise logical OR, and $\mathbb{E}_{\mathbf{M}(t_i)}[\cdot]$ averages over voxels with $\mathbf{M}(t_i)=1$.

\cref{table:ab_perturbation_size} shows that geometry fidelity (CD/HD95) reaches its optimum at $k=1$.
However, it is insensitive to $k$, consistent with time-averaged reconstruction $\overline{\mathbf{V}}_c$ in \cref{fig:ab_perturbation_size}(a).
This is because the averaging process would smooth temporal artifacts or details at individual timestamps that differentiate model performance across different $k$ values.
Therefore, we resort to dynamics-sensitive metrics and observations to determine the optimal $k$.
Indeed, $k=1$ achieves the highest PSNR/SSIM and lowest DLE metrics, corresponding to the best DSA image synthesis quality and dynamics learning accuracy.
Next, we analyze the impact of $k$ \R{on temporal fidelity and small-vessel visibility} using time-resolved contrast volumes $\mathbf{V}_c(t)$ and synthesized DSA images in \cref{fig:ab_perturbation_size}(b,c), focusing on vessels of varying sizes and flow speeds~\citep{stroke}.
\begin{itemize}
    \item \textbf{Large Artery with Fast Flow:}
    Similar to \cref{sec:ab components}, sparse-view overfitting leads to temporally inconsistent floating artifacts without perturbation ($k=0$), as shown in contrast volume at $t_{18}$ (red arrows) and its projection (orange boxes).
    Since these artifacts are high-frequency temporal noise, our perturbation suppresses them as $k$ increases, \R{which improves temporal fidelity.}
    Notably, these large vessels have fast but strong and stable flow, and local contrast concentration changes smoothly over time as a low-frequency signal, thereby avoiding significant reconstruction errors. 

    \item \textbf{Small Vessel with Fast Flow:}
    Typical in rapid collateral circulation~\citep{collateral} (orange boxes at $t_{128}$).
    Rapid blood flow in these small vessels is unstable and fluctuating, leading to abrupt temporal changes in local contrast concentration as a high-frequency signal. 
    Our perturbation struggles to distinguish it from noise.
    Thus, overly large $k$ (e.g., $k=3$) over-smooths it and degrades \R{small-vessel visibility}.

    \item \textbf{Small Vessel with Slow Flow:}
    Typical in distal vasculature~\citep{perfusion-grade}. 
    As shown in contrast volume at $t_{18}$ (blue boxes) and its projection (green boxes), our perturbation enhances \R{small-vessel visibility} by smoothing high-frequency overfitting noise to preserve low-frequency temporal signal from slow-filling blood.
\end{itemize}
These scenarios are most common in our arterial-phase cerebral DSA, with rare cases like slow-flow large arteries excluded.
In all, our choice $k=1$ is optimal for its best trade-off: suppressing artifacts, \R{improving temporal fidelity while} avoiding excessive dynamics errors, and enhancing \R{small-vessel visibility}.

\subsubsection{Hash Encoder Hyperparameters}

\begin{table}[]
\caption{Ablation study on hash encoder hyperparameters.}
\centering
\fontsize{6.5}{6.5}\selectfont
\setlength{\tabcolsep}{7pt}
\renewcommand\arraystretch{1.2}
\begin{tabular}{cccccc}
\toprule
\multicolumn{2}{c}{}              & CD ($\mathrm{mm}$) $\downarrow$ & HD95 ($\mathrm{mm}$) $\downarrow$ & PSNR ($\mathrm{dB}$) $\uparrow$ & SSIM $\uparrow$ \\ \midrule
\multirow{3}{*}{$H$} 
% & $2^{14}$   &  1.36$\pm$0.16                               &  2.44$\pm$0.62                       & 34.44$\pm$1.40                     & 0.830$\pm$0.029  \\
& $2^{15}$   &  1.32$\pm$0.13                               &  2.38$\pm$0.62                       & \textbf{34.46$\pm$1.40}                     & 0.832$\pm$0.029  \\
                           & ${2^{19}}$\textsuperscript{$\dag$}   & \textbf{1.28$\pm$0.15}                    & \textbf{2.33$\pm$0.74}                   & 34.43$\pm$1.35                     & \textbf{0.833$\pm$0.029} \\
                           & $2^{23}$   &  1.31$\pm$0.20                   & 2.60$\pm$1.30                   &          34.38$\pm$1.39       & 0.832$\pm$0.029       \\ \midrule
\multirow{3}{*}{$L$}       & 8    & 1.37$\pm$0.16                   & 2.63$\pm$0.79                  & 34.01$\pm$1.33     & 0.829$\pm$0.028  \\
                           & 12\textsuperscript{$\dag$}   & \textbf{1.28$\pm$0.15}                    & \textbf{2.33$\pm$0.74}                   & \textbf{34.43$\pm$1.35}                     & \textbf{0.833$\pm$0.029} \\
                           & 16   & 1.30$\pm$0.15                     &  2.38$\pm$0.79                  & 34.39$\pm$1.38                         &   0.831$\pm$0.029 \\ \midrule
\multirow{2}{*}{$F$}       & 4    & 1.35$\pm$0.16                   & 2.43$\pm$0.84                  & 34.35$\pm$1.35                      &  \textbf{0.834$\pm$0.036}  \\
                           & 8\textsuperscript{$\dag$}    & \textbf{1.28$\pm$0.15}                    & \textbf{2.33$\pm$0.74}                   & \textbf{34.43$\pm$1.35}                     & 0.833$\pm$0.029 \\ \midrule
\multirow{3}{*}{$b_s, b_p$} & 1.3  & 1.35$\pm$0.17                &         2.64$\pm$0.91       & \textbf{34.56$\pm$1.33}               &    \textbf{0.835$\pm$0.027}     \\
                           & 1.45\textsuperscript{$\dag$} & \textbf{1.28$\pm$0.15}                    & \textbf{2.33$\pm$0.74}                   & 34.43$\pm$1.35                     & 0.833$\pm$0.029 \\
                           & 1.6  & 1.30$\pm$0.15                    & 2.37$\pm$0.79                   &  34.35$\pm$1.36                         &   0.830$\pm$0.029 \\ \midrule
\multirow{3}{*}{$b_d$}     & 1.25 & 1.56$\pm$0.72                   &  2.81$\pm$1.49                  &  33.34$\pm$2.37                  &   0.820$\pm$0.035 \\
                           & 1.4\textsuperscript{$\dag$}  & \textbf{1.28$\pm$0.15}                    & \textbf{2.33$\pm$0.74}                   & 34.43$\pm$1.35                     & \textbf{0.833$\pm$0.029} \\
                           & 1.55 & 1.29$\pm$0.16                  &  2.38$\pm$0.83                &  \textbf{34.48$\pm$1.37}                  & 0.832$\pm$0.029     \\ \bottomrule
\end{tabular}
\label{table:ab_hash_params}
\vspace{-2mm}
\end{table}

We justify the choices of hash encoder hyperparameters, as they can influence model performance.
The quantitative results are provided in \cref{table:ab_hash_params}, where we denote $b_s,b_p$, and $b_d$ as the growth factors of $\bm{h}_s,\bm{h}_p$, and $\bm{h}_d$.
Each row modifies one parameter type while keeping others fixed to default values from \cref{table:hash_params}.
$F$ is capped at 8 (same as Instant-NGP~\citep{Instant-ngp}).
To assess model sensitivity, we vary parameters $L$, $H$, and $F$ simultaneously for all three hash encoders.
Growth factors $b$ are ablated separately per encoder due to their distinct modeling targets.
Results in \cref{table:ab_hash_params} show that while our default setting is not always the best for each individual metric, it consistently yields robust, near-optimal performance across all of them, making it a well-balanced and reliable choice.

Hash encoders’ growth factors are set according to their target modeling resolutions.
The finest hash grid resolution is $R_{L} = \lfloor R_{1} \cdot b^{L-1} \rfloor$.
Specifically, $\bm{h}_s$ and $\bm{h}_p$ are designed to capture spatial information, thus their $R_{L}$ is set to 476 which is close to \R{the} reconstructed volume spatial resolution (512).
In contrast, $\bm{h}_d$ focuses on capturing temporal dynamics, thus its $R_{L}$ is 80 which matches the DSA temporal resolution (133).
All values are set slightly lower due to the sparsity of DSA imaging scene.
Given fixed $L$ and $R_{1}$, these target modeling resolutions \R{determine} our choices for $b_s,b_p$, and $b_d$.
Ablation results in \cref{table:ab_hash_params} further validate that these choices are effective.

\subsection{\R{Robustness Analysis}}

\R{In this section, we further evaluate our method's robustness against real-world acquisition variability including input noise and geometric angular jitter.
Consistent with the ablation study, all experiments are conducted using 40 input views on the WUH dataset.}

\subsubsection{\R{Input Noise}}

\begin{table}[t]
% \color{red}
\caption{\R{Quantitative results of robustness analysis to input noise.
$N_{\text{ph}}$ denotes the incident photon count, where lower values indicate higher noise levels.
`--' denotes noise-free baseline.}}
\centering
\fontsize{6.5}{6.5}\selectfont
\setlength{\tabcolsep}{10pt}
\renewcommand\arraystretch{1.2}
\begin{tabular}{ccccc}
\toprule
$N_{\text{ph}}$ & CD ($\mathrm{mm}$) $\downarrow$ & HD95 ($\mathrm{mm}$) $\downarrow$ & PSNR ($\mathrm{dB}$) $\uparrow$ & SSIM $\uparrow$          \\ \midrule
--        & \textbf{1.28$\pm$0.15}          & \textbf{2.33$\pm$0.74}            & 34.43$\pm$1.35                  & 0.833$\pm$0.029          \\
$10^5$ & 1.34$\pm$0.16                   & 2.44$\pm$0.82                     & \textbf{34.56$\pm$1.38}         & \textbf{0.849$\pm$0.031} \\
$10^4$ & 1.37$\pm$0.20                   & 2.60$\pm$1.20                     & 34.48$\pm$1.38                  & 0.849$\pm$0.032          \\
$10^3$ & 2.23$\pm$0.92                   & 5.93$\pm$5.45                     & 33.58$\pm$1.29                  & 0.809$\pm$0.036          \\ \bottomrule
\end{tabular}
\label{table:rob_noise}
\vspace{-2mm}
\end{table}

\begin{figure}[t]
\centering
    \includegraphics[width=0.48\textwidth]{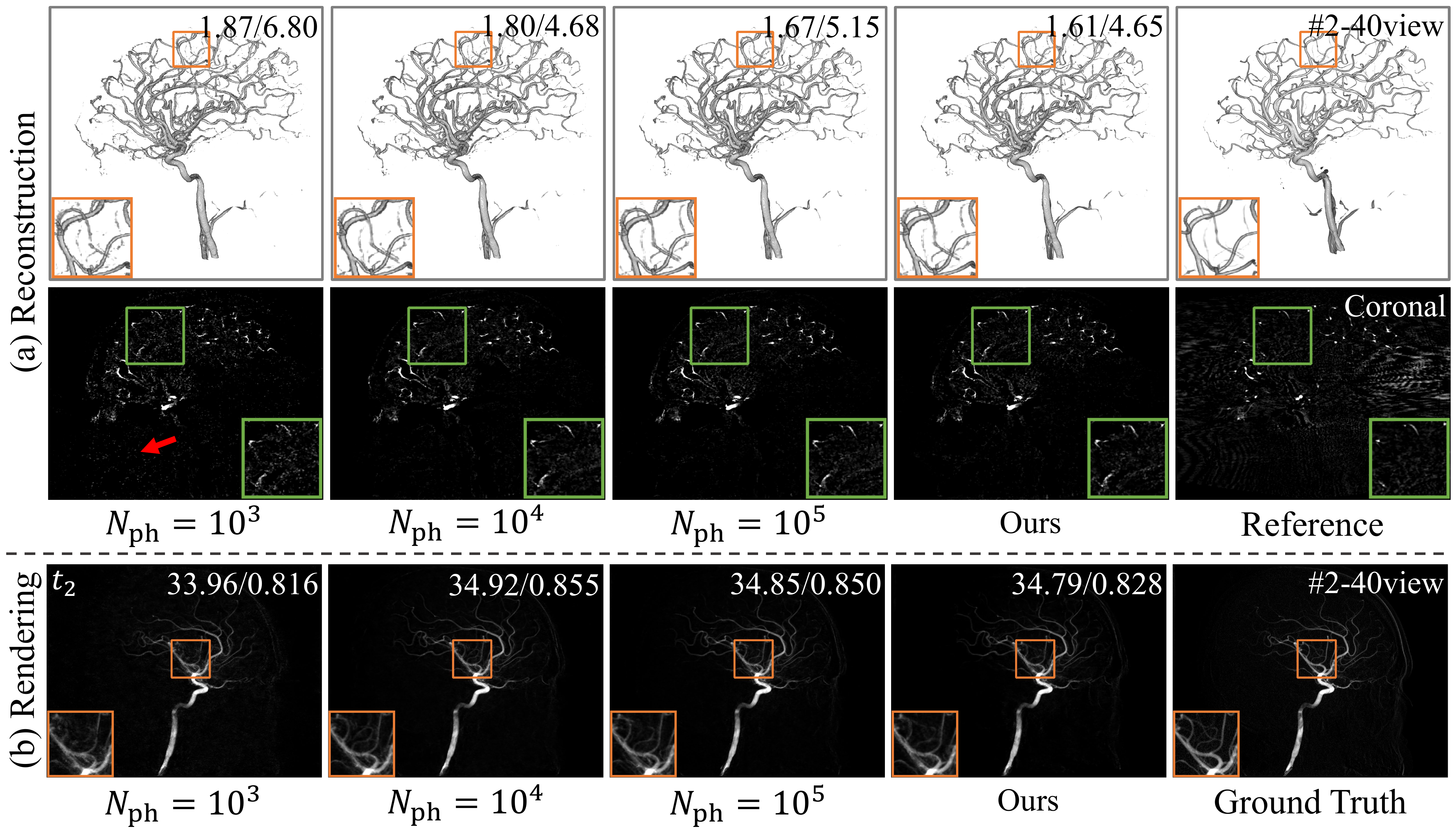}
    \caption{
    \R{Qualitative results of robustness analysis to input noise.
    $N_{\text{ph}}$ denotes the incident photon count, where lower values indicate higher noise levels.
    (a) 3D vessel reconstruction with top-row 3D visualization and bottom-row coronal slice.
    (b) 2D DSA image synthesis at test frame.}
    } 
\label{fig:rob_noise}
\vspace{-2mm}
\end{figure}

\R{We evaluate robustness against input acquisition noise by applying mixed Poisson-Gaussian noise to the training projections using the TIGRE toolbox~\citep{tigre}.
Different X-ray dose levels are simulated by varying the incident photon counts ($N_{\text{ph}} \in \{10^5, 10^4, 10^3\}$), with additive Gaussian electronic noise of fixed standard deviation $\eta = 10$, corresponding to low, moderate, and severe noise levels, respectively.}

\R{As summarized in \cref{fig:rob_noise} and \cref{table:rob_noise}, our model exhibits robustness at low-to-moderate noise levels ($N_{\text{ph}} = 10^5$ and $10^4$).
The reconstruction performance remains stable, with metrics and visual results comparable to the noise-free baseline.
However, performance degrades under the severe noise setting ($N_{\text{ph}} = 10^3$).
This is quantified by deteriorated metrics.
Qualitatively, the 3D reconstruction and synthesized DSA image exhibit mild loss of fine details as marked by orange boxes, while the selected coronal slice reveals increased background noise.
Overall, these findings validate our method's robustness under mild input noise.}

\subsubsection{\R{Geometric Angular Jitter}}

\begin{table}[t]
% \color{red}
\caption{\R{Robustness analysis to geometric angular jitter $\Delta\theta$.
`--' denotes unperturbed baseline.}}
\centering
\fontsize{6.5}{6.5}\selectfont
\setlength{\tabcolsep}{8.5pt}
\renewcommand\arraystretch{1.2}
\begin{tabular}{ccccc}
\toprule
$\Delta\theta$         & CD ($\mathrm{mm}$) $\downarrow$ & HD95 ($\mathrm{mm}$) $\downarrow$ & PSNR ($\mathrm{dB}$) $\uparrow$ & SSIM $\uparrow$          \\ \midrule
--              & \textbf{1.28$\pm$0.15}          & \textbf{2.33$\pm$0.74}            & 34.43$\pm$1.35         & \textbf{0.833$\pm$0.029} \\
$\pm0.1^\circ$ & 1.30$\pm$0.16                   & 2.37$\pm$0.84                     & \textbf{34.45$\pm$1.38}                  &\textbf{0.833$\pm$0.029}          \\
$\pm0.3^\circ$ & 1.32$\pm$0.16                   & 2.49$\pm$0.94                     & 34.31$\pm$1.38                  & 0.831$\pm$0.029          \\
$\pm0.5^\circ$ & 1.36$\pm$0.23                   & 2.72$\pm$1.42                     & 34.04$\pm$1.38                  & 0.827$\pm$0.030          \\ 
$\pm1^\circ$   & 1.56$\pm$0.31                   & 3.62$\pm$2.03                     & 33.24$\pm$1.43                  & 0.816$\pm$0.034          \\ \bottomrule
\end{tabular}
\label{table:rob_angle}
\vspace{-2mm}
\end{table}

\begin{figure}[t]
\centering
    \includegraphics[width=0.48\textwidth]{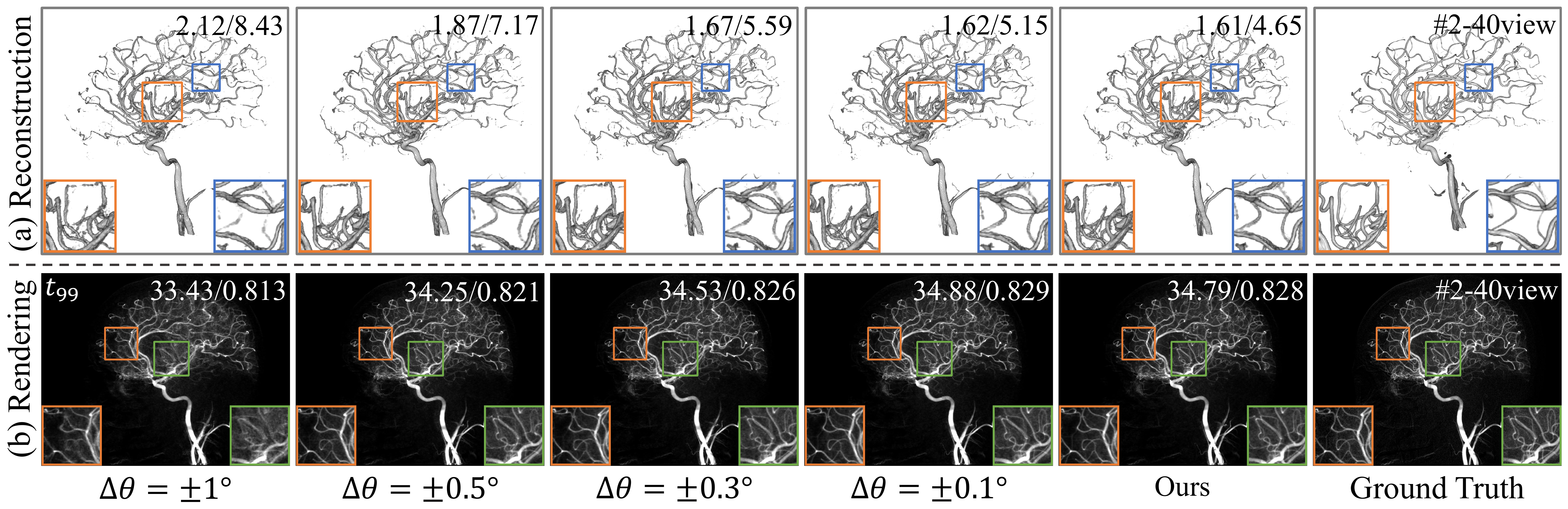}
    \caption{
    \R{Qualitative results of robustness analysis to geometric angular jitter $\Delta\theta$.
    (a) 3D vessel reconstruction.
    (b) 2D DSA synthesis at test frame.}
    } 
\label{fig:rob_angle}
\vspace{-4mm}
\end{figure}

\R{It is important to clarify that cerebral DSA imaging is typically free from significant physiological motion like respiratory or cardiac motion.
Consequently, our assumption of static vessel structure is clinically reasonable.
However, we acknowledge that involuntary patient head motion or system calibration errors may still occur between mask and fill runs.
These factors can lead to geometric inconsistencies as mask-fill misalignment, which may degrade reconstruction quality.}

\R{Since re-acquiring clinical data with controlled motion is infeasible, we evaluate our robustness against such geometric inconsistencies by introducing random angular jitter as a proxy during the training stage.
Specifically, we apply uniform random perturbations to the projection angles for each training view within predefined ranges ($\pm 0.1^\circ, \pm 0.3^\circ, \pm 0.5^\circ, \pm 1^\circ$), while evaluations are performed on the unperturbed geometry.
In our sparse-view setting (40 views), the angular spacing between adjacent training frames is approximately $5^\circ$.
Therefore, a perturbation of $\pm 1^\circ$ represents a relatively severe angular disturbance, corresponding to roughly one-fifth of the sampling interval.}

\R{As summarized in \cref{table:rob_angle} and \cref{fig:rob_angle}, our model maintains robust performance under small-to-moderate perturbations ($\leq 0.5^\circ$).
Both metrics and visual quality show marginal deviations from the unperturbed baseline.
However, severe jitter ($\pm 1^\circ$) leads to performance degradation.
This is characterized by deteriorated metrics and loss of fine vessel details as marked by colored boxes in 3D reconstruction and DSA synthesis.
Overall, these findings validate our method's practical reliability under mild projection angular errors.}

\section{Discussion}
\label{sec:discussion}

Our method achieves high-quality reconstruction from sparse DSA images (e.g., 30 views) in real-world experiments. 
This demonstrates its potential to \R{assist clinical diagnosis, preoperative planning, and hemodynamic analysis while significantly reducing radiation exposure}, thereby safeguarding the health of both patients and radiographers. 
In the following, we discuss our limitations and directions for future research.

\subsection{Clinical Validation and Evaluation Limitations}

Our method was validated on multi-view arterial-phase cerebral DSA data from \R{two institutions, demonstrating its robustness across different clinical centers.}
While this \R{confirms its} effectiveness, broader validation on \R{more} datasets with different scanners and imaging protocols would further establish generalizability in future work.
Moreover, our method is carefully designed for multi-view cerebral DSA scans.
Single-plane and bi-plane acquisitions, common in coronary artery examinations, are beyond our scope due to limited views and unmodeled cardiac motion.
We will explore these scenarios in the future.

% Absolute ground truth for in vivo data is unavailable.
% The scanner-provided reference volume allows surface-based evaluation of 3D vessel geometry, but time-resolved reconstructions are not provided, limiting fully quantitative assessment.
\R{Absolute ground truth for in vivo data is unavailable, and the scanner-provided reference volumes contain inherent imperfections.
To compensate for this, we conducted a blinded expert study to evaluate 3D reconstruction quality and clinical utility.
However, quantitative assessment of time-resolved dynamics remains limited.}
Future work could leverage programmable flow phantoms to enable voxel-wise validation of 3D vessel and 4D blood flow reconstruction, providing a gold standard for quantitative evaluation.

\subsection{Correction for Mask-Fill Misalignment}

\R{Our robustness analysis confirms the method's tolerance to mild projection angular errors.}
\R{However,} for practical clinical deployment, addressing \R{severe} mask-fill misalignment is an important consideration.
A promising solution is to apply 2D image registration as a pre-processing step to correct such misalignment.
This can be achieved via traditional methods (e.g., feature point matching~\citep{GaborDSA}, B-spline registration~\citep{DSA-bspline}), learning-based methods (e.g., VoxelMorph~\citep{voxelmorph}, AngioMoCo~\citep{angiomoco}), or neural field methods like NIR~\citep{NIR}. 
In future work, we will explore such registrations as pre-processing to further enhance clinical robustness.

\subsection{Avenues for Runtime Optimization}
\label{sec:dis_runtimeopt}

\R{As detailed in \cref{sec:efficiency}, our current per-case optimization time ($\sim$2 hours) poses a challenge for time-sensitive acute interventions, such as stroke thrombectomy.
While the proposed method focuses on high-fidelity reconstruction for non-acute workflows including preoperative planning and hemodynamic analysis, improving reconstruction speed is critical for broader clinical adoption.
To this end, we propose two strategies to accelerate training.}

\begin{itemize}
    \item \textbf{Vessel Probability Guided Free Space Skipping:}
    The highly sparse vessel structures in DSA scans create large empty regions.
    Our original volumetric rendering uniformly sampled points along each casting ray, leading to redundant computation in empty regions.
    To improve efficiency, we use our model's time-agnostic vessel probability to identify vessel regions via a coarse occupancy grid ($1/4$ the reconstruction resolution, updated every 16 iterations).
    A grid cell is marked as occupied if its vessel probability exceeds a threshold (0.01), or empty otherwise.
    During rendering, samples in empty space are skipped by checking this grid, avoiding unnecessary network computations.
    \R{This strategy achieves a substantial $3\times$ speed-up, reducing training time from 2 hours to 40 minutes (verified on Case \#1), significantly improving the feasibility for clinical workflows.}    
    It is implemented with Nerfacc toolbox~\citep{nerfacc} using default hyperparameters.

    \item \textbf{Vessel-Aware Training Ray Sampling:}
    Another potential extension is to employ a pretrained vessel segmentation network to guide training ray sampling, with denser sampling in segmented vessel regions instead of our current random sampling across the entire image.
    This strategy may accelerate training convergence and will be explored in future work.

\end{itemize}

\subsection{Potential for Real-Time Inference}
\label{sec:potential_realtime}

Despite high reconstruction fidelity, our method and other per-case optimization approaches require considerable computation time, ranging from several minutes to hours.
Recent advances in large-scale feed-forward architectures enable rapid reconstruction \R{for static 3D scenes} as discussed in \cref{sec:relatedworks_DL}.
However, single-pass inference often yields errors or misses patient-specific fine details, making them unsuitable for direct clinical use.
Therefore, we \R{envision} a more clinically viable two-stage hybrid approach.
\R{First, once we collect a large-scale DSA dataset, a fast feed-forward network could be trained to provide a high-quality initial reconstruction in near real-time.}
Second, efficient optimization-based refinement on the initial result would correct errors and recover fine details, ensuring consistency with input DSA images.
We believe this strategy, which combines feed-forward speed with per-case optimization fidelity, is a crucial step towards real-time clinical DSA reconstruction.
We leave it as a key future direction.

\enlargethispage{1.2\baselineskip}

% \subsection{Broader Applicability}
% \label{sec:discuss broader app}

% Our core paradigm vessel probability guided static-dynamic decomposition is applicable to other cerebral dynamic imaging modalities.
% In Time-Resolved Contrast-Enhanced Magnetic Resonance Angiography (TR CE-MRA)~\citep{TRCEMRA1, TRCEMRA2}, a technique of 4D Magnetic Resonance Imaging (4D MRI), time-independent vessel probability field could similarly separate contrast flow from static non-vascular brain tissue.
% This spatial prior would mitigate spatio-temporal trade-off, enabling high-quality reconstruction from undersampled k-space data.
% For Perfusion CT~\citep{perfusionCT1,perfusionCT2}, a probability field identifying perfused brain parenchyma could decompose contrast perfusion from static non-perfused brain tissue, regularizing reconstruction from sparse or low-dose measurements. 
% Our coarse-to-fine training for stable optimization and temporal perturbed loss for temporal consistency are also applicable to these new contexts and general optimization problems.

\section{Conclusion}

In summary, we propose a NeRF-based optimization framework that successfully \R{addresses} sparse-view DSA reconstruction.
Our method leverages a time-agnostic vessel probability field to guide the attenuation learning, capturing the dynamic nature of DSA imaging.
And we further incorporate two training strategies to improve our model performance: coarse-to-fine progressive training for better vascular geometry and temporal perturbed rendering loss for temporal consistency.
Extensive experiments demonstrate high-quality 3D vessel reconstruction and 2D DSA image synthesis.

\section*{Acknowledgements}
{This work was supported in part by the National Natural Science Foundation of China No. 6230012077, the Shanghai Municipal Central Guided Local Science and Technology Development Fund Project No. YDZX20233100001001, and HPC Platform of ShanghaiTech University.}

\enlargethispage{1.2\baselineskip}

%%Harvard
\bibliographystyle{model2-names.bst}\biboptions{authoryear}
\bibliography{reference}

@article{DSA_app_1,
  title={4D-DSA: development and current neurovascular applications},
  author={Ruedinger, KL and Schafer, S and Speidel, MA and Strother, CM},
  journal={American Journal of Neuroradiology},
  volume={42},
  number={2},
  pages={214--220},
  year={2021},
  publisher={Am Soc Neuroradiology}
}

@article{DSA_app_3,
  title={4D DSA for dynamic visualization of cerebral vasculature: a single-center experience in 26 cases},
  author={Lang, S and G{\"o}litz, P and Struffert, T and R{\"o}sch, J and R{\"o}ssler, K and Kowarschik, M and Strother, C and Doerfler, A},
  journal={American Journal of Neuroradiology},
  volume={38},
  number={6},
  pages={1169--1176},
  year={2017},
  publisher={Am Soc Neuroradiology}
}

@article{NASCET,
  title={How to measure carotid stenosis.},
  author={Fox, Allan J},
  journal={Radiology},
  volume={186},
  number={2},
  pages={316--318},
  year={1993}
}

@article{WASID,
  title={A standardized method for measuring intracranial arterial stenosis},
  author={Samuels, Owen B and Joseph, Gregg J and Lynn, Michael J and Smith, Harriet A and Chimowitz, Marc I},
  journal={American journal of neuroradiology},
  volume={21},
  number={4},
  pages={643--646},
  year={2000},
  publisher={American Journal of Neuroradiology}
}

@article{Demo_to_Neck,
  title={Comparison of 2D digital subtraction angiography and 3D rotational angiography in the evaluation of dome-to-neck ratio},
  author={Brinjikji, W and Cloft, H and Lanzino, G and Kallmes, DF},
  journal={American Journal of Neuroradiology},
  volume={30},
  number={4},
  pages={831--834},
  year={2009},
  publisher={American Journal of Neuroradiology}
}

@inproceedings{DSA_recon_metric,
  title={2D versus 3D comparison of angiographic imaging biomarkers using computational fluid dynamics simulations of contrast injections},
  author={Shields, A and Bhurwani, MMS and Williams, K and Chivukula, V and Bednarek, DR and Rudin, S and Ionita, CN},
  booktitle={Medical Imaging 2023: Physics of Medical Imaging},
  volume={12463},
  pages={480--495},
  year={2023},
  organization={SPIE}
}

@article{FDK,
  title={Practical cone-beam algorithm},
  author={Feldkamp, Lee A and Davis, Lloyd C and Kress, James W},
  journal={Josa a},
  volume={1},
  number={6},
  pages={612--619},
  year={1984},
  publisher={Optica Publishing Group}
}

@article{fdk-3ddsarecon,
  title={Use of a C-arm system to generate true three-dimensional computed rotational angiograms: preliminary in vitro and in vivo results},
  author={Fahrig, R and Fox, AJ and Lownie, S and Holdsworth, DW},
  journal={American Journal of Neuroradiology},
  volume={18},
  number={8},
  pages={1507--1514},
  year={1997},
  publisher={Am Soc Neuroradiology}
}

@article{Single-recon,
  title={Patient-specific reconstruction of volumetric computed tomography images from a single projection view via deep learning},
  author={Shen, Liyue and Zhao, Wei and Xing, Lei},
  journal={Nature biomedical engineering},
  volume={3},
  number={11},
  pages={880--888},
  year={2019},
  publisher={Nature Publishing Group UK London}
}

@inproceedings{X2CTGAN,
  title={X2CT-GAN: reconstructing CT from biplanar X-rays with generative adversarial networks},
  author={Ying, Xingde and Guo, Heng and Ma, Kai and Wu, Jian and Weng, Zhengxin and Zheng, Yefeng},
  booktitle={Proceedings of the IEEE/CVF conference on computer vision and pattern recognition},
  pages={10619--10628},
  year={2019}
}

@article{3DCBCTrecon,
  author={Liu, Zhentao and Fang, Yu and Li, Changjian and Wu, Han and Liu, Yuan and Shen, Dinggang and Cui, Zhiming},
  journal={IEEE Transactions on Medical Imaging}, 
  title={Geometry-Aware Attenuation Learning for Sparse-View CBCT Reconstruction}, 
  year={2025},
  volume={44},
  number={2},
  pages={1083-1097},
  doi={10.1109/TMI.2024.3473970}
}

@inproceedings{DIFnet,
  title="Learning Deep Intensity Field for Extremely Sparse-View CBCT Reconstruction",
  author="Lin, Yiqun and Luo, Zhongjin and Zhao, Wei and Li, Xiaomeng",
  booktitle="Medical Image Computing and Computer Assisted Intervention -- MICCAI 2023",
  pages="13--23",
  year="2023",
  publisher="Springer Nature Switzerland"
}

@article{X-LRM,
  title={X-lrm: X-ray large reconstruction model for extremely sparse-view computed tomography recovery in one second},
  author={Zhang, Guofeng and Zha, Ruyi and He, Hao and Liang, Yixun and Yuille, Alan and Li, Hongdong and Cai, Yuanhao},
  journal={arXiv preprint arXiv:2503.06382},
  year={2025}
}

@article{X-GRM,
  title={X-GRM: Large Gaussian Reconstruction Model for Sparse-view X-rays to Computed Tomography},
  author={Liu, Yifan and Li, Wuyang and Yu, Weihao and Li, Chenxin and Alahi, Alexandre and Meng, Max and Yuan, Yixuan},
  journal={arXiv preprint arXiv:2505.15235},
  year={2025}
}

@article{3DDSArecon,
  title={Self-supervised learning enables 3D digital subtraction angiography reconstruction from ultra-sparse 2D projection views: a multicenter study},
  author={Zhao, Huangxuan and Zhou, Zhenghong and Wu, Feihong and Xiang, Dongqiao and Zhao, Hui and Zhang, Wei and Li, Lin and Li, Zhong and Huang, Jia and Hu, Hongyao and others},
  journal={Cell Reports Medicine},
  volume={3},
  number={10},
  year={2022},
  publisher={Elsevier}
}

@inproceedings{drrref2,
  title={Reconstruction of digital radiographs by texture mapping, ray casting and splatting},
  author={Alakuijala, J and Jaske, UM and Sallinen, S and Hehminen, H and Laitinen, J},
  booktitle={Proceedings of 18th Annual International Conference of the IEEE Engineering in Medicine and Biology Society},
  volume={2},
  pages={643--645},
  year={1996},
  organization={IEEE}
}

@article{NeRF,
  title={Nerf: Representing scenes as neural radiance fields for view synthesis},
  author={Mildenhall, Ben and Srinivasan, Pratul P and Tancik, Matthew and Barron, Jonathan T and Ramamoorthi, Ravi and Ng, Ren},
  journal={Communications of the ACM},
  volume={65},
  number={1},
  pages={99--106},
  year={2021},
  publisher={ACM New York, NY, USA}
}

@inproceedings{Intratomo,
  title={IntraTomo: self-supervised learning-based tomography via sinogram synthesis and prediction},
  author={Zang, Guangming and Idoughi, Ramzi and Li, Rui and Wonka, Peter and Heidrich, Wolfgang},
  booktitle={Proceedings of the IEEE/CVF International Conference on Computer Vision},
  pages={1960--1970},
  year={2021}
}

@inproceedings{NAF,
  title={NAF: neural attenuation fields for sparse-view CBCT reconstruction},
  author={Zha, Ruyi and Zhang, Yanhao and Li, Hongdong},
  booktitle={International Conference on Medical Image Computing and Computer-Assisted Intervention},
  pages={442--452},
  year={2022},
  organization={Springer}
}

@inproceedings{sax_nerf,
  title={Structure-Aware Sparse-View X-ray 3D Reconstruction},
  author={Yuanhao Cai and Jiahao Wang and Alan Yuille and Zongwei Zhou and Angtian Wang},
  booktitle={CVPR},
  year={2024}
}

@article{TiAVox,
  title={TiAVox: Time-aware Attenuation Voxels for Sparse-view 4D DSA Reconstruction},
  author={Zhou, Zhenghong and Zhao, Huangxuan and Fang, Jiemin and Xiang, Dongqiao and Chen, Lei and Wu, Lingxia and Wu, Feihong and Liu, Wenyu and Zheng, Chuansheng and Wang, Xinggang},
  journal={arXiv preprint arXiv:2309.02318},
  year={2023}
}

@inproceedings{DVGO,
  title={Direct voxel grid optimization: Super-fast convergence for radiance fields reconstruction},
  author={Sun, Cheng and Sun, Min and Chen, Hwann-Tzong},
  booktitle={Proceedings of the IEEE/CVF Conference on Computer Vision and Pattern Recognition},
  pages={5459--5469},
  year={2022}
}

@article{Instant-ngp,
  title={Instant neural graphics primitives with a multiresolution hash encoding},
  author={M{\"u}ller, Thomas and Evans, Alex and Schied, Christoph and Keller, Alexander},
  journal={ACM Transactions on Graphics (ToG)},
  volume={41},
  number={4},
  pages={1--15},
  year={2022},
  publisher={ACM New York, NY, USA}
}

@inproceedings{D-nerf,
  title={D-nerf: Neural radiance fields for dynamic scenes},
  author={Pumarola, Albert and Corona, Enric and Pons-Moll, Gerard and Moreno-Noguer, Francesc},
  booktitle={Proceedings of the IEEE/CVF Conference on Computer Vision and Pattern Recognition},
  pages={10318--10327},
  year={2021}
}

@article{neus,
  title={Neus: Learning neural implicit surfaces by volume rendering for multi-view reconstruction},
  author={Wang, Peng and Liu, Lingjie and Liu, Yuan and Theobalt, Christian and Komura, Taku and Wang, Wenping},
  journal={arXiv preprint arXiv:2106.10689},
  year={2021}
}

@inproceedings{neuralangelo,
  title={Neuralangelo: High-Fidelity Neural Surface Reconstruction},
  author={Li, Zhaoshuo and M{\"u}ller, Thomas and Evans, Alex and Taylor, Russell H and Unberath, Mathias and Liu, Ming-Yu and Lin, Chen-Hsuan},
  booktitle={Proceedings of the IEEE/CVF Conference on Computer Vision and Pattern Recognition},
  pages={8456--8465},
  year={2023}
}

@inproceedings{4dhash,
  title={Temporal Interpolation Is All You Need for Dynamic Neural Radiance Fields},
  author={Park, Sungheon and Son, Minjung and Jang, Seokhwan and Ahn, Young Chun and Kim, Ji-Yeon and Kang, Nahyup},
  booktitle={Proceedings of the IEEE/CVF Conference on Computer Vision and Pattern Recognition},
  pages={4212--4221},
  year={2023}
}

@inproceedings{tineuvox,
  title={Fast dynamic radiance fields with time-aware neural voxels},
  author={Fang, Jiemin and Yi, Taoran and Wang, Xinggang and Xie, Lingxi and Zhang, Xiaopeng and Liu, Wenyu and Nie{\ss}ner, Matthias and Tian, Qi},
  booktitle={SIGGRAPH Asia 2022 Conference Papers},
  pages={1--9},
  year={2022}
}

@inproceedings{kplanes,
  title={K-planes: Explicit radiance fields in space, time, and appearance},
  author={Fridovich-Keil, Sara and Meanti, Giacomo and Warburg, Frederik Rahb{\ae}k and Recht, Benjamin and Kanazawa, Angjoo},
  booktitle={Proceedings of the IEEE/CVF Conference on Computer Vision and Pattern Recognition},
  pages={12479--12488},
  year={2023}
}

@article{SART,
  title={Simultaneous algebraic reconstruction technique (SART): a superior implementation of the ART algorithm},
  author={Andersen, Anders H and Kak, Avinash C},
  journal={Ultrasonic imaging},
  volume={6},
  number={1},
  pages={81--94},
  year={1984},
  publisher={SAGE Publications Sage CA: Los Angeles, CA}
}

@article{ASDPOCS,
  title={Image reconstruction in circular cone-beam computed tomography by constrained, total-variation minimization},
  author={Sidky, Emil Y and Pan, Xiaochuan},
  journal={Physics in Medicine \& Biology},
  volume={53},
  number={17},
  pages={4777},
  year={2008},
  publisher={IOP Publishing}
}

@book{beerslaw,
  title={Principles of computerized tomographic imaging},
  author={Kak, Avinash C and Slaney, Malcolm},
  year={2001},
  publisher={SIAM}
}

@inproceedings{Freenerf,
  title={FreeNeRF: Improving Few-shot Neural Rendering with Free Frequency Regularization},
  author={Yang, Jiawei and Pavone, Marco and Wang, Yue},
  booktitle={Proceedings of the IEEE/CVF Conference on Computer Vision and Pattern Recognition},
  pages={8254--8263},
  year={2023}
}

@article{SSIM,
  title={Image quality assessment: from error visibility to structural similarity},
  author={Wang, Zhou and Bovik, Alan C and Sheikh, Hamid R and Simoncelli, Eero P},
  journal={IEEE transactions on image processing},
  volume={13},
  number={4},
  pages={600--612},
  year={2004},
  publisher={IEEE}
}

@ARTICLE{ICP,
  author={Besl, P.J. and McKay, Neil D.},
  journal={IEEE Transactions on Pattern Analysis and Machine Intelligence}, 
  title={A method for registration of 3-D shapes}, 
  year={1992},
  volume={14},
  number={2},
  pages={239-256},
  doi={10.1109/34.121791}
}

@incollection{marchingcube,
  title={Marching cubes: A high resolution 3D surface construction algorithm},
  author={Lorensen, William E and Cline, Harvey E},
  booktitle={Seminal graphics: pioneering efforts that shaped the field},
  pages={347--353},
  year={1998}
}

@article{tigre,
  title={TIGRE: a MATLAB-GPU toolbox for CBCT image reconstruction},
  author={Biguri, Ander and Dosanjh, Manjit and Hancock, Steven and Soleimani, Manuchehr},
  journal={Biomedical Physics \& Engineering Express},
  volume={2},
  number={5},
  pages={055010},
  year={2016},
  publisher={IOP Publishing}
}

@article{nerfacc,
  title={NerfAcc: Efficient Sampling Accelerates NeRFs.},
  author={Li, Ruilong and Gao, Hang and Tancik, Matthew and Kanazawa, Angjoo},
  journal={arXiv preprint arXiv:2305.04966},
  year={2023}
}

@article{Open3D,
  title={Open3D: A modern library for 3D data processing},
  author={Zhou, Qian-Yi and Park, Jaesik and Koltun, Vladlen},
  journal={arXiv preprint arXiv:1801.09847},
  year={2018}
}

@article{3D_slicer,
  title={3D Slicer as an image computing platform for the Quantitative Imaging Network},
  author={Fedorov, Andriy and Beichel, Reinhard and Kalpathy-Cramer, Jayashree and Finet, Julien and Fillion-Robin, Jean-Christophe and Pujol, Sonia and Bauer, Christian and Jennings, Dominique and Fennessy, Fiona and Sonka, Milan and others},
  journal={Magnetic resonance imaging},
  volume={30},
  number={9},
  pages={1323--1341},
  year={2012},
  publisher={Elsevier}
}

@inproceedings{focalloss,
  title={Focal loss for dense object detection},
  author={Lin, Tsung-Yi and Goyal, Priya and Girshick, Ross and He, Kaiming and Doll{\'a}r, Piotr},
  booktitle={Proceedings of the IEEE international conference on computer vision},
  pages={2980--2988},
  year={2017}
}

@inproceedings{mlpbias,
  title={On the spectral bias of neural networks},
  author={Rahaman, Nasim and Baratin, Aristide and Arpit, Devansh and Draxler, Felix and Lin, Min and Hamprecht, Fred and Bengio, Yoshua and Courville, Aaron},
  booktitle={International conference on machine learning},
  pages={5301--5310},
  year={2019},
  organization={PMLR}
}

@article{LRM,
  title={Lrm: Large reconstruction model for single image to 3d},
  author={Hong, Yicong and Zhang, Kai and Gu, Jiuxiang and Bi, Sai and Zhou, Yang and Liu, Difan and Liu, Feng and Sunkavalli, Kalyan and Bui, Trung and Tan, Hao},
  journal={arXiv preprint arXiv:2311.04400},
  year={2023}
}

@inproceedings{GRM,
  title={Grm: Large gaussian reconstruction model for efficient 3d reconstruction and generation},
  author={Xu, Yinghao and Shi, Zifan and Yifan, Wang and Chen, Hansheng and Yang, Ceyuan and Peng, Sida and Shen, Yujun and Wetzstein, Gordon},
  booktitle={European Conference on Computer Vision},
  pages={1--20},
  year={2024},
  organization={Springer}
}

@inproceedings{VGGT,
  title={Vggt: Visual geometry grounded transformer},
  author={Wang, Jianyuan and Chen, Minghao and Karaev, Nikita and Vedaldi, Andrea and Rupprecht, Christian and Novotny, David},
  booktitle={Proceedings of the Computer Vision and Pattern Recognition Conference},
  pages={5294--5306},
  year={2025}
}

@article{DSA-bspline,
  title={Nonrigid image registration in digital subtraction angiography using multilevel B-Spline},
  author={Nejati, Mansour and Sadri, Saeid and Amirfattahi, Rassoul},
  journal={BioMed research international},
  volume={2013},
  number={1},
  pages={236315},
  year={2013},
  publisher={Wiley Online Library}
}

@article{voxelmorph,
  title={Voxelmorph: a learning framework for deformable medical image registration},
  author={Balakrishnan, Guha and Zhao, Amy and Sabuncu, Mert R and Guttag, John and Dalca, Adrian V},
  journal={IEEE transactions on medical imaging},
  volume={38},
  number={8},
  pages={1788--1800},
  year={2019},
  publisher={IEEE}
}

@article{NIR,
  title={Medical image registration via neural fields},
  author={Sun, Shanlin and Han, Kun and You, Chenyu and Tang, Hao and Kong, Deying and Naushad, Junayed and Yan, Xiangyi and Ma, Haoyu and Khosravi, Pooya and Duncan, James S and others},
  journal={Medical Image Analysis},
  volume={97},
  pages={103249},
  year={2024},
  publisher={Elsevier}
}

@inproceedings{angiomoco,
  title={Angiomoco: Learning-based motion correction in cerebral digital subtraction angiography},
  author={Su, Ruisheng and van der Sluijs, Matthijs and Cornelissen, Sandra and van Zwam, Wim and van der Lugt, Aad and Niessen, Wiro and Ruijters, Danny and van Walsum, Theo and Dalca, Adrian},
  booktitle={International Conference on Medical Image Computing and Computer-Assisted Intervention},
  pages={770--780},
  year={2023},
  organization={Springer}
}

@inproceedings{GaborDSA,
  title={DSA image registration based on multiscale Gabor filters and mutual information},
  author={Cao, Zhiguo and Liu, Xiaoxiao and Peng, Bo and Moon, Yiu-Sang},
  booktitle={2005 IEEE International Conference on Information Acquisition},
  pages={6--pp},
  year={2005},
  organization={IEEE}
}

@book{stroke,
  title={Stroke e-book: Pathophysiology, diagnosis, and management},
  author={Grotta, James C and Albers, Gregory W and Broderick, Joseph P and Kasner, Scott E and Lo, Eng H and Sacco, Ralph L and Wong, Lawrence KS and Day, Arthur L},
  year={2021},
  publisher={Elsevier Health Sciences}
}

@article{collateral,
  title={Time-resolved assessment of collateral flow using 4D CT angiography in large-vessel occlusion stroke},
  author={Fr{\"o}lich, Andreas MJ and Wolff, Sarah Lena and Psychogios, Marios N and Klotz, Ernst and Schramm, Ramona and Wasser, Katrin and Knauth, Michael and Schramm, Peter},
  journal={European radiology},
  volume={24},
  number={2},
  pages={390--396},
  year={2014},
  publisher={Springer}
}

@article{perfusion-grade,
  title={Recommendations on angiographic revascularization grading standards for acute ischemic stroke: a consensus statement},
  author={Zaidat, Osama O and Yoo, Albert J and Khatri, Pooja and Tomsick, Thomas A and Von Kummer, R{\"u}diger and Saver, Jeffrey L and Marks, Michael P and Prabhakaran, Shyam and Kallmes, David F and Fitzsimmons, Brian-Fred M and others},
  journal={Stroke},
  volume={44},
  number={9},
  pages={2650--2663},
  year={2013},
  publisher={Lippincott Williams \& Wilkins Hagerstown, MD}
}

@article{hypernerf,
  title={Hypernerf: A higher-dimensional representation for topologically varying neural radiance fields},
  author={Park, Keunhong and Sinha, Utkarsh and Hedman, Peter and Barron, Jonathan T and Bouaziz, Sofien and Goldman, Dan B and Martin-Brualla, Ricardo and Seitz, Steven M},
  journal={arXiv preprint arXiv:2106.13228},
  year={2021}
}

@article{Emernerf,
  title={Emernerf: Emergent spatial-temporal scene decomposition via self-supervision},
  author={Yang, Jiawei and Ivanovic, Boris and Litany, Or and Weng, Xinshuo and Kim, Seung Wook and Li, Boyi and Che, Tong and Xu, Danfei and Fidler, Sanja and Pavone, Marco and others},
  journal={arXiv preprint arXiv:2311.02077},
  year={2023}
}

@inproceedings{nerfies,
  title={Nerfies: Deformable neural radiance fields},
  author={Park, Keunhong and Sinha, Utkarsh and Barron, Jonathan T and Bouaziz, Sofien and Goldman, Dan B and Seitz, Steven M and Martin-Brualla, Ricardo},
  booktitle={Proceedings of the IEEE/CVF international conference on computer vision},
  pages={5865--5874},
  year={2021}
}

@inproceedings{HexPlane,
  title={Hexplane: A fast representation for dynamic scenes},
  author={Cao, Ang and Johnson, Justin},
  booktitle={Proceedings of the IEEE/CVF Conference on Computer Vision and Pattern Recognition},
  pages={130--141},
  year={2023}
}

@inproceedings{Dynamic_nerf,
  title={Dynamic view synthesis from dynamic monocular video},
  author={Gao, Chen and Saraf, Ayush and Kopf, Johannes and Huang, Jia-Bin},
  booktitle={Proceedings of the IEEE/CVF International Conference on Computer Vision},
  pages={5712--5721},
  year={2021}
}

@article{masked_hash,
  title={Masked space-time hash encoding for efficient dynamic scene reconstruction},
  author={Wang, Feng and Chen, Zilong and Wang, Guokang and Song, Yafei and Liu, Huaping},
  journal={Advances in neural information processing systems},
  volume={36},
  pages={70497--70510},
  year={2023}
}

\end{document}